\documentclass{modified}
\usepackage{bm}
\usepackage{bbm}
\usepackage{enumitem}
\usepackage[dvipsnames]{xcolor}

\pdfoutput=1

\newcommand{\eps}{\epsilon}
\newcommand{\defeq}{\overset{\text{def}}{=}}

\newcommand{\bilform}[2]{\mathcal{B}(#1,#2)}

\lefttitle{L. R. Seitz and B. A. Wingate}

\title{Quasi-geostrophic limiting dynamics and energetics of the LANS-$\alpha$ model}

\author{L. R. Seitz\aff{1} and Beth A. Wingate\aff{2}}

\affiliation{\aff{1} Division of Applied Mathematics, Brown University, Providence, RI 02906, USA
\aff{2} Department of Mathematics, Harrison Building, University of Exeter, Exeter EX4 4PY, UK}

\corresau{L. R. Seitz, \texttt{lrseitz@brown.edu}}

\begin{document}
\maketitle

\begin{abstract}
The Lagrangian-Averaged Navier-Stokes-$\alpha$ (LANS-$\alpha$) model, a turbulence closure scheme based on energy-conserving modifications to nonlinear advection, can produce more energetic simulations than standard models, leading to improved fidelity (e.g., in ocean models). However, comprehensive understanding of the mechanism driving this energetic enhancement has proven elusive. To address this gap, we derive the fast quasi-geostrophic limit of the three-dimensional, stably-stratified LANS-$\alpha$ equations. This provides both the slow, balanced flow and the leading-order fast wave dynamics. Analysis of these wave dynamics suggests that an explanation for the energetic enhancement lies in the dual role of the smoothing parameter itself: increasing $\alpha$ regularizes the dynamics and simultaneously generates a robust landscape of wave-wave resonant interactions. Direct numerical simulations show that $\alpha$ plays an analogous role to that of the Burger number ($Bu$) in governing the partition of energy between slow and fast modes -- and consequently, the timescale of geostrophic adjustment -- but with key differences. Increasing $\alpha$, regardless of the relative strengths of rotation and stratification, extends the lifetime of wave energy by delaying the dominance of the slow modes. We find that the creation of an energy pathway only involving fast waves is a universal outcome of the regularization across all values of $Bu$, contrasting with a disruption of slow-fast interactions that is most impactful only in the $Bu=1$ case. These insights unify the LANS-$\alpha$ model's characteristic energetic enhancement with, in some cases, its known numerical stiffness, identifying potential pathways to mitigate stability issues hindering the broader application of LANS-$\alpha$-type models.
\end{abstract}


\section{Introduction}\label{sec:introduction}

\par The Lagrangian-Averaged Navier-Stokes-$\alpha$ (LANS$-\alpha$) equations constitute a turbulence closure model that has achieved at lower resolution what other more commonly used closure schemes require higher resolution to accomplish, in a number of contexts \citep{chen1999direct, holm2005review, zhao2005dynamic, geurts-holm, hecht2008lans, hecht2008}. Notably, LANS-$\alpha$ produces more energetic simulations \citep{holmnadiga2003, petersen2008}. This improves heat and salinity transport resolution \citep{hecht2008lans} and turbulence statistics \citep{bennis2021lans} in ocean models and representations of eddy momentum flux in atmosphere models \citep{aizinger2015}. The $\alpha$-regularization also lowers the critical wavenumber for baroclinic instability \citep{holmwingate2005baroclinic}, improving its ability to model the transfer of potential to kinetic energy on coarser grids. 
\par The LANS-$\alpha$ model is fundamentally different from classical closure schemes: it regularizes nonlinear advection rather than introducing an eddy diffusivity. This functions to filter the nonlinear dynamics at length scales smaller than the namesake parameter $\alpha$. The effectiveness of the LANS-$\alpha$ model, as well as its key physical and mathematical properties --- such as a Kelvin circulation theorem \citep{holmtiti2005}, a K\'arm\'an-Howarth theorem \citep{holm2002karman}, a conserved energy \citep{holmfluctuations}, bounded enstrophy \citep{farhatjollylunasin2014}, well-posedness \citep{marsden2001}, and the existence of a finite-dimensional global attractor \citep{kim2006, kim2018} --- are perhaps unsurprising, given that its formulation emerged independently from several lines of inquiry. For instance, while its mathematical roots trace back to Leray's seminal work, in which regularizing Navier-Stokes is the key idea needed to prove the existence of weak solutions \citep{leray1934}, it more directly originates from the Gjaja-Holm Wave-Mean Flow Interaction (WMFI) equations \citep{gjajaholm1996}. The LANS-$\alpha$ equations also share mathematical connections with closure schemes that parameterize eddy stress based on deformation \citep{portamanazanna2014, anstey2017deformation, bachman2018}. As in those methods, numerical instabilities can sometimes arise when implementing the LANS-$\alpha$ model, limiting its broader application \citep{hecht2008, petersen2008, bachman2018}. Both the significant advantages and the primary obstacle of the LANS-$\alpha$
model lead to the central question of this work: can the model's energetic enhancement be understood through the parameter $\alpha$'s role in modifying the underlying wave dynamics and energy transfers?
\begin{figure}
    \centering
    \includegraphics[width=\linewidth]{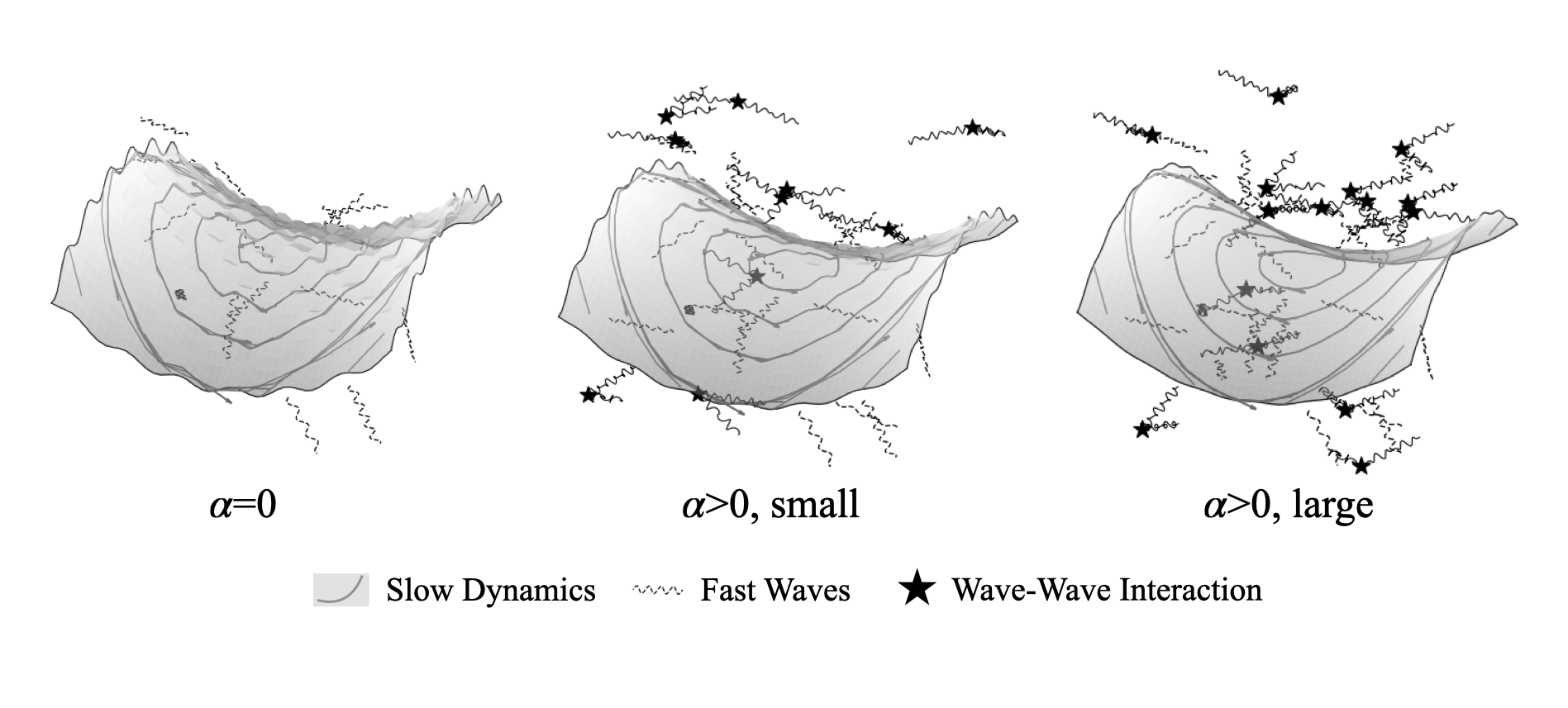}
    \caption{A schematic diagram of the slow and fast dynamics within the fast QG-$\alpha$ limit, for different values of the regularization parameter $\alpha$. The slow dynamics are illustrated by grey streamlines (over a grey background) and are smoother as $\alpha$ increases, reflecting the role $\alpha$ is found to have in the three-dimensional QG-$\alpha$ equations \eqref{eq:qg-alpha-v2}. The fast waves are shown as the dashed black lines distinct from the slow dynamics, and their interaction is symbolized by a star. Although there are fast waves in the $\alpha=0$ case, for many Burger numbers, there are no interactions between fast waves that contribute to the dynamics until $\alpha>0$. The number of interactions increases with $\alpha$, as shown in $\S$\ref{subsec:fast-dynamics}. This modification of the energetic landscape by $\alpha$ may help to explain why the LANS-$\alpha$ model is more energetic than other models.}
    \label{fig:lans-schematic}
\end{figure}
\par While there are partial explanations for why the LANS-$\alpha$ model is more energetic --- varying $\alpha$ modifies the dissipation range \citep{chen1999direct}, a property shared by related $\alpha$-regularization models \citep{ilyinlunasintiti2006, lunasin2008}, and introduces an effective Rossby deformation radius, enabling baroclinic instability to be resolved at coarser resolution \citep{holmwingate2005baroclinic} --- further insight into the processes driving this increased energy would benefit both the theoretical understanding of the model and its implementation. To this end, we investigate the energetics of the LANS-$\alpha$ model within the quasi-geostrophic (QG) limit. 
\par The QG equations, originally derived by \cite{charney1948}, describe the large-scale, balanced dynamics primarily driven by planetary rotation. The three-dimensional, continuously stratified system considered here is further constrained by strong background stratification. By design, the QG equations effectively filter out fast, small-scale motions like inertia-gravity waves, isolating the dominant ``balanced" flow. The first derivation of the QG limiting dynamics for the Boussinesq-$\alpha$ equations appears in \citep{HolmMarsdenRatiu2002_EPE_GFD}. We extend \citep{HolmMarsdenRatiu2002_EPE_GFD} by presenting the first fast singular limit derivation of QG-$\alpha$ for the Boussinesq-$\alpha$ model. In our work, the leading-order asymptotic solution is the conservation of PV-$\alpha$ modulated by the leading-order fast dynamics, which is found by deriving an operator to project the full dynamics onto the null space of the fast operator. Using this framework, we show there exists a rich dynamical structure involving interactions purely among fast waves (fast-fast-fast dynamics) that is absent in the unregularized ($\alpha = 0$) case (Figure \ref{fig:lans-schematic}). Further, the prognostic variable PV-$\alpha$ derived in this limit is intrinsically useful for future studies of LANS-$\alpha$; it is the central dynamical variable of the balanced system and a quantity fundamental to understanding large-scale geophysical flows (e.g., \citealt{charney1948}; \citealt{rhinesyoung1982}; \citealt{hoskins1985}; \citealt{vallis1996}; \citealt{holmfluctuations}; \citealt{julienknobloch2006}). In other words, we derive QG-$\alpha$ by formulating a projection operator onto the balanced flow. This projection serves a dual purpose: it also provides the foundation for our primary goal of analyzing the energetics of LANS-$\alpha$ by enabling the partitioning of the system's energy between the balanced and fast wave dynamics.
\par The complete elimination of fast dynamics without affecting the evolution of the mean flow, as is intended in Charney's QG equations, has been described as the existence of a ``slow manifold," a concept furthered by \cite{leith1980} and \cite{lorenz1980}. Ideally, a reduced equation resulting from an asymptotic expansion in physical parameters of interest would correspond to a slow manifold: attracting other dynamics toward balanced states, representing dominant dynamics in a lower-dimensional subspace, and ensuring that trajectories initialized on this subspace remain there. In reality, this ideal is rarely achieved, and the ``slow manifolds'' that can be obtained are better described as ``fuzzy" slow manifolds \citep{warnmenard1986}. Even if an original set of equations permits a corresponding slow manifold in the mathematical sense, it may not serve as the desired physical slow manifold, e.g., in the QG limit, one where gravity wave activity is entirely absent \citep{lorenz1987nonexistence, jacobs1991existence, lorenz1992slow}. Indeed, the QG equations do not represent a perfectly invariant subspace, since even perfectly balanced initial conditions can generate exponentially small inertia-gravity waves \citep{vannesteyavneh2004}. Furthermore, even though fast and slow dynamics are decoupled at leading order, they are connected in subtle but dynamically significant ways.  Resonant interactions allow slow (also called potential vorticity or geostrophic) modes to affect the distribution of fast waves \citep{warddewar2010}. When forcing is not restricted to the slow dynamics, a common scenario in realistic settings, fast dynamics can significantly impact the distribution of energy throughout the system \citep{smith2002,smithlee2005,whiteheadwingate2014}. A robust analysis of the energetics of the LANS-$\alpha$ model thus requires not only deriving the slow QG-$\alpha$ equations but also accounting for the influence of the fast dynamics. 
\par To do so, we depart from methods that strictly isolate a slow manifold and instead take a fast singular limit \citep{klainerman1981, majda1984, babin1995, em1996, em1998,em1998a}.  By taking the same parameter limit as traditional QG -- low Rossby number ($Ro\to 0$, geostrophic balance) and low Froude number ($Fr\to 0$, hydrostatic balance), while the ratio between them ($F= Fr/Ro=f/N$) remains fixed -- this approach recovers the standard QG equations for the slow dynamics while retaining fast dynamics at leading order. Applying this process to the LANS-$\alpha$ equations yields a system we term the ``fast QG-$\alpha$ dynamics" (or fast QG-$\alpha$ limit). The fast QG-$\alpha$ dynamics decompose into the slow dynamics, the balanced geostrophic flow corresponding to the traditional QG equations (but with $\alpha$ regularization), and the fast dynamics, the inertia-gravity waves filtered out in the traditional limit. The fast dynamics are decoupled from the usual slow dynamics in the sense that the leading-order fast dynamics cannot influence the slow dynamics because these fast, oscillatory effects cancel out when averaged across many wave periods. In this context, the fluctuations about the balanced QG-$\alpha$ state are the $O(1)$ fast dynamics together with all $O(\eps)$ dynamics, where the small parameter $\eps$ is taken to be the Rossby number. Physically, the frequency of the fast-wave oscillations is inversely proportional to $\eps$. This approach \citep{em1996, em1998} is rigorously justified by Schochet's \citeyearpar{schochet1994} method of cancellation of oscillations for hyperbolic PDEs and may be thought of as \emph{averaging} over fast waves in geophysical flows.  
\par The three-dimensional Boussinesq (or in this work, Boussinesq-$\alpha$) equations admit two kinds of eigenfrequencies: slow (geostrophic) modes (vortical modes) which have zero frequency for all wavenumbers $\bm{k}$ and fast (ageostrophic) modes (dispersive waves). While potential vorticity in the fast QG (or QG-$\alpha$) limit is determined solely by slow modes, in other balanced limits such as low Rossby/finite Froude or low Froude/finite Rossby, waves associated with the slower physical process can contribute to the slow dynamics \citep{em1998, babin1997, wingate2011}. 
\par To understand what drives the LANS-$\alpha$ model's distinctive energetics, we analyze three-wave resonant and near-resonant interactions. Resonant interactions between slow and fast modes, i.e., interactions between geostrophic turbulence and waves, are a primary mechanism for geostrophic adjustment via a direct cascade of wave energy \citep{bartello1995}. These resonant interactions are highly dependent on the Burger number ($Bu=Ro^2/Fr^2$), which is related to the ratio $F=f/N$ via $Bu=1/F^2$. For example, \cite{babin1997} show analytically that as stratification increases compared to rotation, ageostrophic energy cascades are ``unfrozen," allowing nonlinear geostrophic adjustment to occur. With small-scale forcing, geostrophic dynamics dominate at large scales for $1/2\leq F\leq 2$ (or equivalently, $1/4\leq Bu\leq 4$), a range with no triad interactions exclusively involving fast waves \citep{smith2002}. With large-scale forcing, the system transitions from being vortical-energy dominated for $F\leq 1$ ($Bu\geq 1$) to being wave-energy dominated for $F>1$ ($Bu<1$) \citep{sukhatmesmith2008}. Near-resonant interactions, while more challenging to analyze \citep{babin1999} and weak for $F\sim O(1)$ \citep{babin2002}, also play an important role in the transfer of energy to slow wave modes \citep{smithlee2005}.   
\par Our numerical simulations reveal that the smoothing parameter $\alpha$ plays a role analogous to that of the rotation-to-stratification ratio $F$ and affects wave-vortex energy ratios in a similar manner. However, a key difference emerges when varying $\alpha$ as opposed to $F$ (or $Bu$): increasing $\alpha$ markedly extends the persistence of wave energy, thereby slowing the typical transition to vortical energy dominance. A likely mechanism for this is found in how $\alpha$ reshapes three-wave resonant interactions. In the standard fast QG limit, resonant interactions in which two fast waves produce another fast wave are nonexistent for many Burger numbers and sparse otherwise. In contrast, by numerically tabulating how often our analytically-derived eigenfrequencies satisfy the condition for resonance interactions, we find that $\alpha$-regularization drastically \emph{increases} the number of these interactions (as illustrated in Figure \ref{fig:lans-schematic}). While $\alpha$ was previously known to modify the dissipation range, these alterations to the triad interactions reveal that the introduction of $\alpha$ can both cause energy to accumulate within the waves and decrease resonant interactions generally responsible for cross-scale energy transfer, providing a mechanism by which energy may be prevented from reaching dissipative scales and thus accumulate within the system. Collectively, our results indicate that LANS-$\alpha$ is likely more energetic than other models because of the way $\alpha$-regularization modulates fast waves, affecting their persistence and ability to satisfy resonance conditions that drive energy transfer among the slow and fast components of the flow.
\par The remainder of this paper is organized as follows. In $\S 2$, we introduce the LANS-$\alpha$ equations, including their non-dimensional, non-local form, and interpretation as Lagrangian-averaged Navier-Stokes. The derivation of the fast QG limit for LANS-$\alpha$ is presented in $\S 3$; this section includes energy conservation laws, the slow dynamics (conservation of PV-$\alpha$), the fast limiting dynamics ($O(1)$ wave dynamics), and an analysis of how the parameter $\alpha$ modifies three-wave resonance interactions. In $\S 4$, we present numerical simulations focusing on the evolution of slow and wave energies in the QG limit and how it depends on $\alpha$. Finally, in $\S 5$, we discuss the main findings and their broader implications.

\section{The LANS-\texorpdfstring{$\alpha$}{alpha} equations}
\subsection{Non-local form of the LANS-\texorpdfstring{$\alpha$}{alpha} equations}
The LANS$-\alpha$ equations with the Boussinesq approximation (the Boussinesq-$\alpha$ equations) are given by \citep{holmfluctuations,holmtiti2005}: 
\begin{subequations}\label{eq:lans-alpha-eqs}
\begin{align}
    \frac{\partial \bm{v}}{\partial t} + \bm{u} \cdot \nabla \bm{v} + \bm{v}\cdot \nabla \bm{u}^T + f\hat{\bm{z}}\times \bm{u} + \nabla \phi+\frac{\rho}{\rho_0}g\hat{\bm{z}} &= \nu \nabla^{2n} \bm{v}, \label{eq:lans-momentum} \\
    \frac{\partial \rho}{\partial t} + \bm{u} \cdot \nabla \rho -\bm{u}\cdot b\hat{\bm{z}} &=\kappa \nabla^{2n} \rho, \label{eq:lans-buoyancy}  \\
    \nabla \cdot \bm{u} &= 0, \\ 
    \bm{v} &= \bm{u} - \alpha^2 \Delta \bm{u}, \label{eq:smoothed-velocity}
\end{align} 
\end{subequations}
where $\bm{v}=(v_1,v_2,v_3)$ is the three-dimensional velocity, $f$ is the Coriolis parameter, $\phi$ is a modified pressure, $\rho$ is the buoyancy variable (density), $\rho_0$ is a reference density, $g$ is the gravitational constant, $\nu$ is kinematic viscosity, $b$ is the strength of the underlying density stratification, $\kappa$ is buoyancy diffusivity, and $\alpha\in \mathbb{R}$ is a parameter with units of length. While this length scale $\alpha$ can be interpreted in many ways -- such as the typical deviation of a particle trajectory from its time-averaged path \citep{holm2005review} or a filter width \citep{lunasin2007} -- we adopt the interpretation that it is the length scale below which the dynamics are regularized. This regularization is accomplished through the introduction of a new velocity field, $\bm{u}$, as defined in \eqref{eq:smoothed-velocity}. In this framework, the smoothed field $\bm{u}$ advects the momentum $\bm{v}$, mitigating the development of sharp gradients at scales smaller than $\alpha$.
\par The difference between the Boussinesq-$\alpha$ equations and the usual Boussinesq equations is essentially due to the introduction of the smoothed velocity $\bm{u}=(u_1, u_2, u_3)$. Denoting the Helmholtz operator as 
\begin{equation}\label{def:helmholtz-operator}
\mathcal{S}\defeq (1-\alpha^2\Delta )
\end{equation}
($\mathcal{S}$ for ``smoothing"), the smoothed velocity $\bm{u}$ is constructed via $\mathcal{S}^{-1}\bm{v}$ \eqref{eq:smoothed-velocity}. The smoothed velocity contributes an additional advective term, $\bm{v} \cdot \nabla \bm{u}^T$ (also written as $(\nabla \bm{u})^T \bm{v}$ or $v_j\partial_i u_j$), in \eqref{eq:lans-momentum} and necessitates a modification to the pressure. The usual pressure $p$ has been replaced with the modified pressure $\phi$ in \eqref{eq:lans-momentum}, where 
\begin{equation}\label{eq:modified-pressure}
\phi \defeq \frac{p}{\rho_0} - \frac{1}{2}|\bm{u}|^2-\frac{\alpha^2}{2}|\nabla \bm{u}|^2.
\end{equation}
From the definition of $\mathcal{S}$ \eqref{def:helmholtz-operator}, the incompressibility of $\bm{u}$ implies that of $\bm{v}$, and vice versa, since the order of the divergence and the Laplacian can be interchanged. Note also that $\mathcal{S}$ can be applied to scalar functions, with the interpretation that $\Delta$ is the scalar Laplacian rather than the vector Laplacian. Using the Helmholtz operator to smooth the advecting velocity does not introduce dissipative effects, since \eqref{eq:lans-momentum} and \eqref{eq:lans-buoyancy} are time-reversible when $\nu=\kappa=0$.
\par In \eqref{eq:lans-momentum}-\eqref{eq:smoothed-velocity}, the density has been decomposed as 
\begin{equation}
\rho_{\text{full}} = \overline{\rho}+\rho, \quad \overline{\rho} = p_0 - bz, \quad b>0.
\end{equation}
\par The dissipation terms in \eqref{eq:lans-momentum} are presented in a general form, and when $n=1$, the dissipation is of the usual Laplacian type. The domain is $[0,L]^3$, for a length scale $L$, with triply periodic boundary conditions, to facilitate the comparison of the results presented here to those of previous turbulence studies and the classical fast singular QG limit \citep{em1998}.
\par We  non-dimensionalize the Boussinesq-$\alpha$ equations by considering a characteristic length scale $L$ (e.g., the domain length) and velocity scale $U$, which correspond to the advective time scale $L/U$. Using the Brunt-V\"ais\"al\"a frequency $N=(gb/\rho_0)^{1/2}$, we define the buoyancy fluctuation scale $B=NU/g$. Note that throughout this work, $N$ is considered constant. We non-dimensionalize $\bm{x}'=\bm{x}/L$, $t'=t/T$, $\bm{v'}=\bm{v}/U$, $\rho'=\rho/(\rho_0 B)$, and the pressure via $p'=p/\overline{p}$ where $\overline{p}$ is a reference pressure (c.f. \citealt{majda-textbook}). Using these scalings, the Rossby, Froude, Reynolds, Prandtl, and Euler numbers arise naturally:
\begin{equation}
Ro \defeq \frac{U}{fL}, \quad Fr \defeq \frac{U}{NL}, \quad  Re \defeq \frac{\rho_0UL}{\nu}, \quad Pr \defeq \frac{\nu}{\rho_0\kappa}, \quad Eu \defeq \frac{\overline{p}}{\rho_0 U^2}.
\end{equation}
The parameter $F$, which measures the relative strength of rotation versus stratification,
\begin{equation}\label{def:F}
    F\defeq \frac{Fr}{Ro}=\frac{f}{N},
\end{equation}
also arises in the system and must be assumed fixed when taking the QG limit. This parameter is closely related to the Burger number 
\begin{equation}\label{def:Bu}
    Bu\defeq \frac{Ro^2}{Fr^2}=\frac{1}{F^2},
\end{equation}
and will play an important role in the dynamics.
\par Omitting the primes, the non-dimensionalized version of \eqref{eq:lans-alpha-eqs} is given by 
\begin{subequations}\label{eq:lans-alpha-eqs-nondim}
\begin{align}
\frac{\partial \bm{v}}{\partial t} + \bm{u}\cdot \nabla \bm{v} +\bm{v} \cdot \nabla \bm{u}^T+ \frac{1}{Ro} \hat{\bm{z}} \times \bm{u} + \nabla \phi + \Gamma\rho\hat{\bm{z}} &= \frac{1}{Re}\nabla^{2n} \bm{v},\label{eq:lans-nondim-momentum} \\
\frac{\partial \rho}{\partial t} + \bm{u}\cdot\nabla\rho - \frac{1}{Fr}\bm{u}\cdot\hat{\bm{z}}  &= \frac{1}{RePr}\nabla^{2n} \rho, \label{eq:lans-nondim-buoyancy}\\ 
\nabla \cdot \bm{u} &= 0, \label{eq:lans-nondim-incompressibility} \\
\bm{u}&=\mathcal{S}^{-1}\bm{v}.
\end{align}
\end{subequations}
Here, $\mathcal{S}$ is actually the non-dimensionalized Helmholtz operator, $(1-L_\alpha^2 \Delta)$, where $L_\alpha \defeq \alpha/L$. Since $\alpha$ is a fraction of the domain length, and we use the domain size as the characteristic length scale, $L_\alpha\in [0,1]$. The modified pressure $\phi$ is also the non-dimensional version. 
The constant $\Gamma\defeq BgLU^{-2}$ but we will use the scaling $\Gamma = Fr^{-1}$. We will consider $n=1$ in the diffusivity for simplicity. 
\par We obtain the non-local form of \eqref{eq:lans-alpha-eqs-nondim} by solving the elliptic equation for the modified pressure, so that
\begin{align}\label{eq:lans-nondim-2}
\nabla \phi = \nabla\Delta^{-1}\left(\frac{1}{Ro}\hat{\bm{z}}\cdot \mathcal{S}^{-1} \bm{\omega} -\nabla\cdot (\mathcal{S}^{-1}\bm{v}\cdot \nabla \bm{v} + \bm{v} \cdot \nabla (\mathcal{S}^{-1}\bm{v})^T)-\frac{1}{Fr}\rho\bm{\hat{z}}\right).
\end{align}
The modified pressure thus does not further differentiate the Bousinessq-$\alpha$ equations from the unregularized Boussinesq equations; $\phi$ is a free variable in exactly the same sense $p$ is. 
\par This system reduces to the usual Boussinesq equations in the case of $\alpha=0$, so we expect the slow and fast dynamics derived for the LANS-$\alpha$ model to reduce to those found in \cite{em1998}, \cite{babin1997}, and \cite{wingate2011} upon setting $\alpha=0$.
\subsection{Interpretation as Lagrangian-Averaged Navier-Stokes}
\par The definition of the smoothed velocity \eqref{eq:smoothed-velocity} indicates that $\bm{u}$ should be interpreted as the Lagrangian average of $\bm{v}$ to justify that this system is ``Lagrangian-averaged" Navier-Stokes. It must then be the case that the Helmholtz operator, applied to a Lagrangian-averaged field, yields an Eulerian-averaged field, and vice versa with the inverse. When defined as the Helmholtz operator dependent on the scalar $\alpha$, $\mathcal{S}$ is an approximation to the true operator for which this is (asymptotically) true. This operator, the dynamical Helmholtz operator, is given by $(1-\tilde{\Delta})$ where $\tilde{\Delta} = \nabla \cdot \langle \boldsymbol{\xi}\boldsymbol{\xi}\rangle \cdot \nabla$ \citep{holmfluctuations}. Here, $\boldsymbol{\xi}$ represents a displacement fluctuation of a particle trajectory $\bm{X}$, so that $\langle \boldsymbol{\xi}\boldsymbol{\xi}\rangle$ is the covariance of the displacement fluctuation. It can be seen to $o(|\boldsymbol{\xi}|^2)$ that the quantity 
\begin{equation}
(\nabla \cdot\langle \boldsymbol{\xi}\boldsymbol{\xi}\rangle \cdot \nabla) \langle \bm{U}\rangle^L,
\end{equation}
where $\langle \bm{U}\rangle^L$ is a Lagrangian mean velocity, is the Stokes drift. Thus, the dynamical Helmholtz operator $(1-\tilde{\Delta})$ can be interpreted as mapping Lagrangian-averaged fields to their Eulerian-averaged counterparts. 
\par However, the LANS-$\alpha$ equations do not involve the \emph{dynamical} Helmholtz operator; $\tilde{\Delta}$ is replaced with $\alpha^2 \Delta$. This assumes the isotropy condition $\langle \xi^k \xi^l\rangle = \alpha^2 \delta^{kl}$, where $\alpha$ is constant if the statistics are homogeneous. Thus, by assuming statistically stationary and isotropic turbulence (in a sense), LANS-$\alpha$ can be interpreted as Lagrangian-averaged Navier-Stokes. This interpretation allows the dynamical Helmholtz operator to be commuted with the advective operator, a key property used in the derivation the slow dynamics. While it is by now standard to regularize Navier-Stokes by applying a number of different filters, this particular filter, with the given meaning of $\alpha$, is especially advantageous due to its commutative properties with space and time derivatives \emph{and} advection. 

\section{The fast quasi-geostrophic limit of the LANS-\texorpdfstring{$\alpha$}{alpha} equations}
\subsection{Abstract framework} 
As in \cite{em1996, em1998}, \cite{babin1997}, \cite{wingate2011}, and \cite{whiteheadwingate2014}, we write the dependent variables as a vector, 
\begin{equation}
    \bm{w} \defeq \begin{pmatrix} \bm{v} \\ \rho\end{pmatrix}.
\end{equation}
The non-dimensionalized LANS-$\alpha$ equations \eqref{eq:lans-alpha-eqs-nondim} may then be rewritten in the abstract operator form 
\begin{equation}\label{eq:abstract-operator-form}
    \frac{\partial \bm{w}}{\partial t} + \frac{1}{Ro}\mathcal{L}_{Ro}\bm{w} + \frac{1}{Fr}\mathcal{L}_{Fr}\bm{w} + \mathcal{B}(\bm{w},\bm{w}) = \mathcal{D}\bm{w}.
\end{equation} 
In \eqref{eq:abstract-operator-form}, the operators are defined:
\begin{equation}\label{eq:lans-abstract-operators}
\begin{split}
    \mathcal{L}_{Ro} \bm{w} &=  \begin{pmatrix}\mathcal{S}^{-1}\Big(\bm{v}_H^\perp+
    \nabla_H \Delta^{-1} \left(\hat{z}\cdot(\nabla \times \bm{v})\right)\Big) \\ \mathcal{S}^{-1}\Big(\frac{\partial}{\partial z}\Delta^{-1}(\hat{z}\cdot(\nabla \times \bm{v}))\Big) \\ 0 
 \end{pmatrix}, \\
 \mathcal{L}_{Fr}\bm{w} &= \begin{pmatrix}   - \nabla_H \Delta^{-1}\left(\frac{\partial \rho}{\partial z}\right)  \\ -\Delta^{-1}\left(\frac{\partial^2\rho}{\partial z^2} \right)+ \rho \\ -\mathcal{S}^{-1}w \end{pmatrix},\\
    \mathcal{B}(\bm{w}, \bm{w}) &= \begin{pmatrix} (\mathcal{S}^{-1} \bm{v} \cdot \nabla) \bm{v} - \nabla \Delta^{-1}\Big(\nabla \cdot ((\mathcal{S}^{-1} \bm{v} \cdot \nabla) \bm{v}+ \bm{v} \cdot \nabla (\mathcal{S}^{-1}\bm{v})^T)\Big) \\ (\mathcal{S}^{-1}\bm{v}\cdot \nabla) \rho \end{pmatrix}, \\ 
    \mathcal{D}\bm{w} &= \frac{1}{Re}\begin{pmatrix} \Delta \bm{v} \\ \frac{1}{Pr}\Delta \rho\end{pmatrix}.
\end{split}
\end{equation}
To write the operators in this form, we made use of the commutative properties of $\mathcal{S}$. Each of these operators reduces to those in \cite{em1998}, \cite{babin1997}, and \cite{wingate2011} in the case of $\alpha=0$.
\par We decompose the full solution to \eqref{eq:abstract-operator-form}, $\bm{w}$, as 
\begin{equation}\label{eq:slow-fast-decomp}
\bm{w} = \bm{w}_F+\bm{w}_S,
\end{equation}
where the subscript $F$ is for ``fast" and the subscript $S$ is for ``slow."
This decomposition is possible because there exists a projection operator $P$ onto the null space of the fast operator (see Appendix \ref{app:slow-fast-decomp}), in a triply periodic domain, such that 
\begin{equation}\label{eq:fast-slow-decomp-projection}
    P\bm{w}_S = \bm{w}_S, \quad P\bm{w}_F = 0.
\end{equation}
In the fast QG-$\alpha$ limit, the fast operator is given by $F\mathcal{L}_{Ro}+\mathcal{L}_{Fr}$. We will use the decomposition \eqref{eq:fast-slow-decomp-projection} to find evolution equations for the components of the flow and energy on and off the slow manifold in the proceeding sections.
\par Both the derivation of the slow dynamics and the fast dynamics involves a spectral analysis of the fast operator; for simplicity, we consider \eqref{eq:abstract-operator-form} with triply periodic boundary conditions.
\subsection{Energy conservation laws} 
The equation for the global integrated total energy is given by (see Appendix \ref{appC}) 
\begin{equation}\label{eq:total-energy-conservation}
    \frac{1}{2}\frac{d}{dt} \int_\Omega \Big(\bm{u}\cdot \bm{v} + \rho^2\Big) \,\text{d}\bm{x}= -\frac{1}{Re}\int_\Omega (\nabla\times \bm{u})\cdot(\nabla \times \bm{v}) - \frac{1}{RePr}(\nabla \rho \cdot \nabla\rho) \,\text{d}\bm{x}.
\end{equation}
\par Since the LANS-$\alpha$ equations can be written in the abstract operator form \eqref{eq:abstract-operator-form} and the fast operator is skew-Hermitian, the results of \cite{em1998} and \citet{babin1995, babin1996, babin1996b, babin1997a, babin1997} together with the energy conservation equation \eqref{eq:total-energy-conservation} yield that 
\begin{equation}\label{eq:energy-decomposition}
    \bm{u}(0)\cdot\bm{v}(0) + \rho(0)^2 = \bm{v}_F\cdot \bm{u}_F + \rho_F^2 + \bm{v}_S\cdot\bm{u}_S+\rho_S^2,
\end{equation}
where in the notation of \eqref{eq:slow-fast-decomp}, $\bm{w}_F=(\bm{v}_F,\rho_F)^T$, $\bm{w}_S=(\bm{v}_S,\rho_S)^T$, and an analogous decomposition for $\mathcal{S}^{-1}\bm{w}$ defines $\bm{u}_F$ and $\bm{u}_S$. Thus the ratio of \emph{total} (as in, kinetic plus potential) fast and slow energy is conserved in the limit as $\eps \to 0$ ($\eps=Ro$). 
\par This allows the energy of the slow balanced flow and that of the fast waves to be treated as distinct quantities. The energy partition is necessary to investigate the mechanism by which the Lagrangian-averaging parameter $\alpha$ alters the interplay between slow vortical and fast wave modes.
\subsection{Slow dynamics, PV-\texorpdfstring{$\alpha$}{alpha}}\label{subsec:slow-dynamics}
To find the slow dynamics, we formulate a projection operator onto the null space of the fast linear operator, as in \eqref{eq:fast-slow-decomp-projection}. A projection is given by (see Appendix \ref{app:deriv-proj-op})
\begin{equation}\label{qg-proj-alpha-neq0}
    P_{\text{QG}_\alpha}\bm{w} = \begin{pmatrix} 
    \bm{v}_H - F^2\mathcal{S}^{-2} \Delta_{\text{QG}_\alpha}^{-1}\frac{\partial^2 \bm{v}_H}{\partial z^2} - \Delta_{\text{QG}_\alpha}^{-1}\left(\nabla_H(\nabla_H\cdot \bm{v}_H) + F\mathcal{S}^{-1} \nabla_H^\perp\left(\frac{\partial \rho}{\partial z}\right)\right)\\ 0 \\ \rho -F \mathcal{S}^{-1}\Delta_{\text{QG}_\alpha}^{-1}\left(\frac{\partial}{\partial z}(\nabla_H\times \bm{v}_H)\right) - \Delta^{-1}_{\text{QG}_\alpha}\Delta_H\rho\end{pmatrix}, 
\end{equation}
where $\Delta_{\text{QG}_\alpha} = \Delta_H + F^2\mathcal{S}^{-2}\frac{\partial^2}{\partial z^2}$, and $\mathcal{S}^{-2}$ denotes $\mathcal{S}^{-1}$ applied twice. When $\alpha=0$, since $\mathcal{S}=\mathcal{S}^{-1}=I$, this correctly reduces to the projection operator in the QG, $\alpha=0$ case in \cite{whiteheadwingate2014}.
\par The (slow) QG-$\alpha$ dynamics are (see Appendix \ref{app:qg-pv-alpha-deriv})
\begin{equation}\label{eq:qg-alpha-v1}
\frac{D^{H,\alpha}}{Dt}\left(\Delta_H\mathcal{S} \phi + F^2 \mathcal{S}^{-1}\frac{\partial^2}{\partial z^2}\phi\right)=0,
\end{equation}
where $\frac{D^{H,\alpha}}{Dt}\defeq \frac{\partial}{\partial t} + \bm{u}_H\cdot\nabla$, and $\phi$ denotes the potential (here, the modified pressure), so that the vertical vorticity $\omega_3= \mathcal{S} \Delta_H \phi$. 
By using the commutative properties of $\mathcal{S}$, we can see that analogously to the $\alpha=0$ case, the potential vorticity is simply the $\Delta_{\text{QG}_\alpha}$ operator applied to the potential $\phi$. We can also rewrite \eqref{eq:qg-alpha-v1} as
\begin{equation}\label{eq:qg-alpha-v2}
\frac{D^{H,\alpha}}{Dt}\left(\omega_3 - F \mathcal{S}^{-1}\frac{\partial}{\partial z}\rho\right)=0,
\end{equation}
which is very similar to the PV conservation equation in the $\alpha=0$ case, 
\begin{equation}\label{eq:qg-alphazero}
    \frac{D^H}{Dt}\left(\omega_3 - F \frac{\partial}{\partial z}\rho\right)=0,
\end{equation}
where the vertical vorticity reduces to $\omega_3 =\Delta_H p$ as usual, since $\mathcal{S}=I$ in the $\alpha=0$ case. The key differences between (the slow portion of) QG-$\alpha$, \eqref{eq:qg-alpha-v2}, and the usual QG, \eqref{eq:qg-alphazero}, are that in QG-$\alpha$, the density is now smoothed, the vorticity has a different relationship to the (modified) pressure, now involving the Helmholtz operator, and the material derivative is with respect to the smoothed velocity. Thus, the dominant, slow QG-$\alpha$ dynamics are regularized compared to the dominant, slow QG $(\alpha=0)$ dynamics (Figure \ref{fig:lans-schematic}): an expected outcome given that LANS-$\alpha$ is a regularized model. 

\subsection{Fast limiting dynamics and modifications of triad interactions by \texorpdfstring{$\alpha$}{alpha}}\label{subsec:fast-dynamics}
While $\alpha$ regularizes the dynamics, we find that it alters the balance between the slow and fast dynamics, amplifying the role of the fast dynamics. To show this, we analyze the triad interactions within the LANS-$\alpha$ equations and compare to those of the unregularized equations. The dispersion relations for the modes associated with the fast QG-$\alpha$ limit are found by seeking eigenfunctions of the fast operator $\mathcal{L}_F$ of the form 
\begin{equation}\label{def:eigenfunctions}
    \bm{w} = \exp(i\bm{k}\cdot \bm{x} - i\omega(\bm{k})t)\bm{r},
\end{equation}
where $\bm{r}$ is an associated eigenvector, $\omega(\bm{k})$ is an associated eigenfrequency, and $\bm{k}=(k_1,k_2,k_3)$ is a wavenumber. In this context, $\omega(\bm{k})$ will be either zero or the classic inertia-gravity wave dispersion relation. We also denote 
\begin{equation}\label{def:wavenumber-mag}
    |\bm{k}|\defeq (k_1^2+k_2^2+k_3^2)^{\frac{1}{2}} \quad \text{and} \quad |\bm{k}_H|\defeq(k_1^2+k_2^2)^{\frac{1}{2}}.
\end{equation}
Analytic formulas for the eigenfrequencies are required to characterize all possible three-wave resonances. In the usual fast QG limit, the eigenfrequencies are given by \citep{em1998}:
\begin{equation}\label{eq:eigenvalues-QG}
   \omega^{(0)}(\bm{k})=0 \quad \text{(double root)} \quad \text{ and } \quad \omega^{(\pm 1)}(\bm{k}) = \pm \frac{(|\bm{k}_H|^2+F^2k_3^2)^{\frac{1}{2}}}{|\bm{k}|}. 
\end{equation}
Physically, the zero frequency (slow) modes are zero frequency Rossby waves, while the modes with frequency $\omega^{(\pm 1)}$ are (dispersive) inertia-gravity waves. The parameter $F=f/N$ appears in the dispersion relation \eqref{eq:eigenvalues-QG} and sets the relative importance of vertical versus horizontal structure. For instance, when stratification is stronger than rotation ($N>f)$, the impact of vertical variations on the inertia-gravity wave frequency is reduced.
\par In the fast QG-$\alpha$ limit, the eigenfrequencies are modified by the non-dimensional Helmholtz operator in spectral space, $(1+L_\alpha|\bm{k}|^2)$, where $L_\alpha=\alpha/L$ is the non-dimensionalized version of $\alpha$. The eigenfrequencies when $\alpha \neq 0$ ($L_\alpha \neq 0$) are given by
\begin{equation}\label{eq:eigenvalues-QG-alpha}
   \omega^{(0)}(\bm{k})=0 \quad \text{(double root)} \quad \text{ and } \quad \omega^{(\pm 1)}(\bm{k}) = \pm \frac{((1+L_\alpha|\bm{k}|^2)|\bm{k}_H|^2+F^2k_3^2)^{\frac{1}{2}}}{(1+L_\alpha|\bm{k}|^2)|\bm{k}|}.
\end{equation}
The smaller $L_\alpha$ is, the closer the $\alpha$-regularized dynamics are to the unregularized dynamics. When $L_\alpha=0$, the $\alpha$-regularized dynamics reduce to the usual dynamics; notice that \eqref{eq:eigenvalues-QG-alpha} reduces to \eqref{eq:eigenvalues-QG} when $L_\alpha=0$. 
\begin{table}

\begin{tabular}{p{2.1cm}p{3.0cm}p{4.0cm}p{3.5cm}}

Resonance type & $\alpha=0$ condition& $\alpha\neq 0$ condition & Role in energy evolution \\
\hline
Slow-Slow-Slow & All wavenumbers & All wavenumbers & Slow only; \par upscale and downscale \par transfer\\ \\
Slow-Fast-Fast &  $F=1$: \par all wavenumbers \par  $F \neq 1$: \par $\left|\frac{k_3}{|\bm{k}_H|}\right|\approx \left|\frac{q_3}{|\bm{q}_H|}\right|$& $F=1$: \par $|\bm{k}|\approx |\bm{q}|$ \par $F \neq 1$: \par $\left|\frac{k_3}{(1+L_\alpha|\bm{k}|^2)|\bm{k}_h|}\right|\approx \left|\frac{q_3}{(1+L_\alpha|\bm{q}|^2)|\bm{q}_H|}\right|$ & Fast only, leaves slow mode unchanged; important for downscale transfer in QG limit\\ \\
Fast-Fast-Fast & None for 
\par $\frac{1}{2}\leq F \leq 2$, \par sparse otherwise & None for $$\frac{1}{2} \leq \frac{1+L_\alpha k_3^2}{F(1+L_\alpha|\bm{k}_h|^2)^{\frac{1}{2}}}\leq 2,$$ possible when $\frac{1}{2}\leq F\leq 2$ and less sparse otherwise  & Fast only   \\ \\
Slow-Slow-Fast & No exact resonances; most efficient fast-slow exchanges associated with near resonances where $\mathbf{k} \leq \mathbf{p} \approx \mathbf{q}$ & Still no exact resonances, but location of near-resonances may be distorted by the spectral Helmholtz operator & Fast and slow; the only type that can produce fast-slow energy exchanges\\ \hline
\end{tabular}
\centering
\caption{Summary of the four types of relevant resonant and near-resonant interactions. Note that Fourier coefficients for fast-fast-slow interactions are zero \citep{em1998}, which does not change when $\alpha \neq 0$. For wavenumbers $\bm{k}+\bm{p}=\bm{q}$, the eigenfrequency condition for (all but slow-slow-slow) resonances to occur changes with the parameter $\alpha$. The resonance types are listed in order of their frequency and importance for energy transfer in the case $\alpha=0$. Slow-slow-slow interactions are the most numerous but are unaffected by $\alpha$. Fast-fast-fast interactions have been less considered because they occur only sparsely, though this changes when $\alpha \neq 0$. For more details, see Appendix \ref{appB2}.}
\label{tab:resonances}
\end{table}
\par The set of resonant triads is described by 
\begin{equation}\label{def:resonant-set}
R_{\beta, \bm{q}}^\alpha \defeq \{(\bm{k}, \bm{p}, \beta', \beta'')\;|\; \bm{k}+\bm{p}=\bm{q}, \; \omega^{(\beta')}(\bm{k})+\omega^{(\beta'')}(\bm{p}) = \omega^{(\beta)}(\bm{q})\},
\end{equation}
where the superscript $\alpha$ emphasizes that the set of possible interactions now depends on $\alpha$ due to \eqref{eq:eigenvalues-QG-alpha}. The relevant classes of triads are described in Table \ref{tab:resonances}. Of these, we consider the slow-fast-fast and fast-fast-fast triads, as the set of possible slow-slow-slow resonances does not change with $\alpha$ and there are no exact slow-slow-fast resonances both when $\alpha=0$ and $\alpha\neq 0$.
\par Slow-fast-fast interactions, when $F\approx 1$ ($Bu\approx 1$), have been seen to result in a rapid downscale energy transfer, and have been described as catalytic interactions responsible for the emergence of a geostrophically adjusted state \citep{bartello1995}. When the eigenfruencies are given by \eqref{eq:eigenvalues-QG}, the condition for wavenumbers $\bm{k}$ and $\bm{p}$ such that $\bm{k}+\bm{p}=\bm{q}$ to be an element of $R_{\beta,\bm{q}}^0$ (with $\beta'=\pm 1, \beta''=0, \beta=\pm 1$), as defined in \eqref{def:resonant-set}, is
\begin{equation}\label{eq:slow-fast-fast-cond-alphazero}
    \pm \frac{(|\bm{k}_H|^2+F^2k_3^2)^{\frac{1}{2}}}{|\bm{k}|} = \pm  \frac{(|\bm{q}_H|^2+F^2q_3^2)^{\frac{1}{2}}}{|\bm{q}|},
\end{equation}
or similarly with $\bm{p}$ in place of $\bm{k}$, interchanging the roles of $\beta'$ and $\beta''$. The condition \eqref{eq:slow-fast-fast-cond-alphazero} produces a resonance for any $\bm{k}, \bm{p}, \bm{q}$ when $F=1$ or a near-resonance when $F=O(1)$. When the eigenfrequencies are instead given by \eqref{eq:eigenvalues-QG-alpha}, the condition is 
\begin{equation}\label{eq:slow-fast-fast-cond-alpha}
    \pm \frac{((1+L_\alpha|\bm{k}|^2)|\bm{k}_H|^2+F^2 k_3^2)^{\frac{1}{2}}}{(1+L_\alpha|\bm{k}|^2)|\bm{k}|} =  \pm \frac{((1+L_\alpha|\bm{q}|^2)|\bm{q}_H|^2+F^2 q_3^2)^{\frac{1}{2}}}{(1+L_\alpha|\bm{q}|^2)|\bm{q}|},
\end{equation}
or similarly with $\bm{p}$ upon interchanging the roles of $\beta'$ and $\beta''$. 

\par
\begin{figure}[t]
    \centering
    \includegraphics[width=0.8\linewidth]{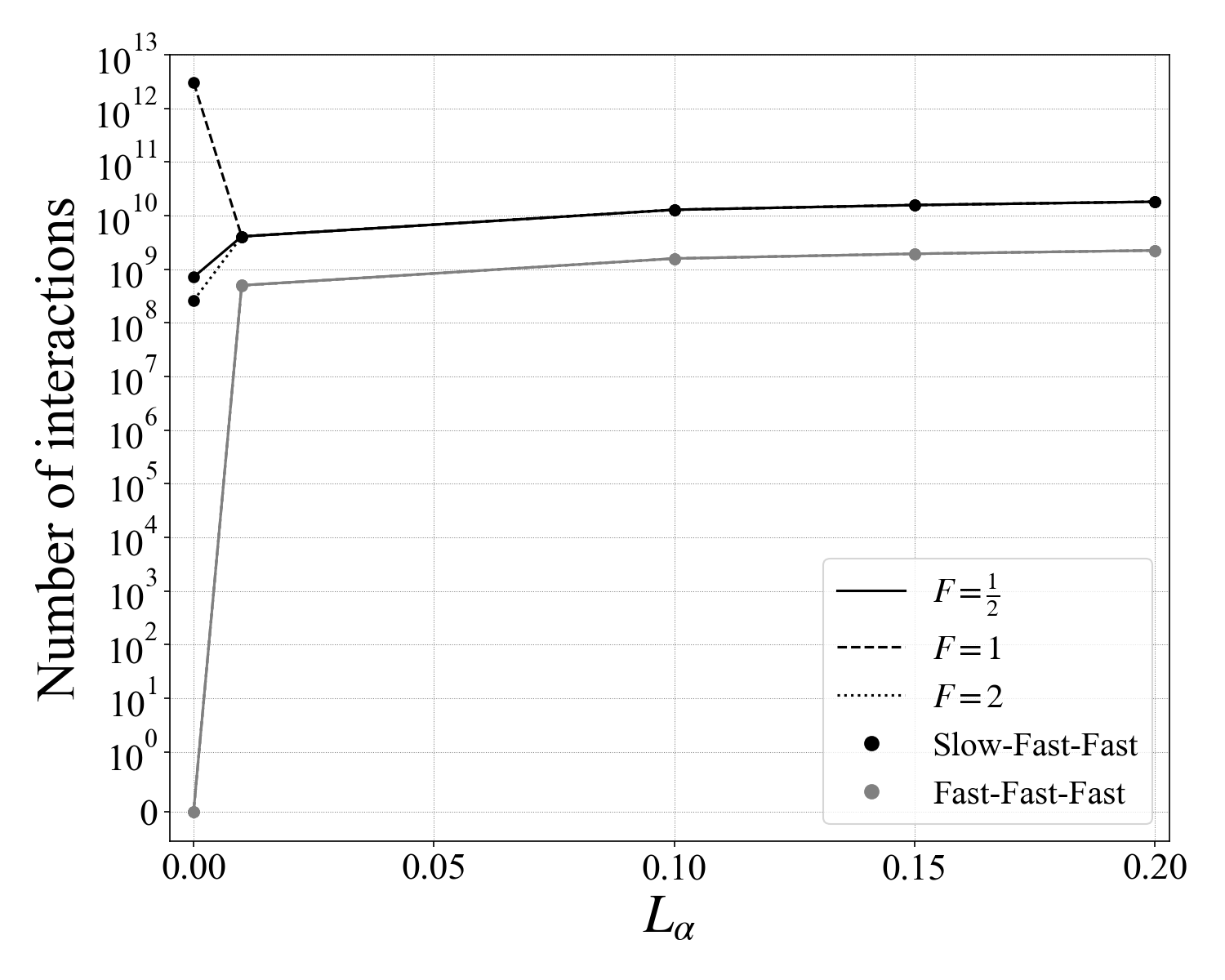}
    \caption{Regularization by $\alpha$ ($L_\alpha=\alpha/L$) reshapes the landscape of resonant interactions, superseding the ratio $F=f/N$ (thus also the Burger number, $Bu=N^2/f^2$) as the dominant parameter determining the relative amounts of different resonances, once $L_\alpha$ is sufficiently large. As $L_\alpha$ increases, the counts of both slow-fast-fast and fast-fast-fast resonant and near-resonant interactions converge to values largely independent of $F$. All counts were computed on a grid of $256^3$ using a tolerance of $10^{-5}$ for numerically checking the resonance conditions. Given the large number wavenumbers associated with the $256^3$ grid, it was possible for the resonance conditions to be satisfied a large number of times (up to trillions, as shown here). To focus on general three-dimensional interactions, triads involving purely horizontal or purely vertical wavenumbers were excluded from the count.}
    \label{fig:2}
\end{figure}

 Due to the additional nonlinearity of \eqref{eq:slow-fast-fast-cond-alpha} as compared to \eqref{eq:slow-fast-fast-cond-alphazero}, it is challenging to reason analytically exactly how the number and distribution of the elements of $R_{\beta,\bm{q}}^\alpha$ changes with $\alpha$ (or $L_\alpha$), but it is possible to do so numerically. Our results show that for $F=1$ ($Bu=1$), the introduction of a nonzero $\alpha$ drastically curtails the number of slow-fast-fast resonant triads, including the cross-scale interactions responsible for transferring energy between large-scale and small-scale dynamics (Figure \ref{fig:2}). This reduction disrupts the primary channel for downscale energy transfer. Further increasing $L_\alpha$ leads to an increase in the number of resonances, though it does not return to the number when $L_\alpha=0$ for reasonable values of $L_\alpha$ (given that in modeling studies, typically $\alpha$ is only a few grid points in length, i.e., $\alpha = 2\Delta x$). For the regimes in which this energy pathway is already less active, $F=1/2$ ($Bu=4$) and $F=2$ ($Bu=1/4$), the introduction of nonzero $\alpha$ has a different effect, only increasing the number of resonances and thus altering rather than shutting down energy exchange via slow-fast-fast resonances. 
\begin{figure}[t]
    \centering
    \includegraphics[width=\linewidth]{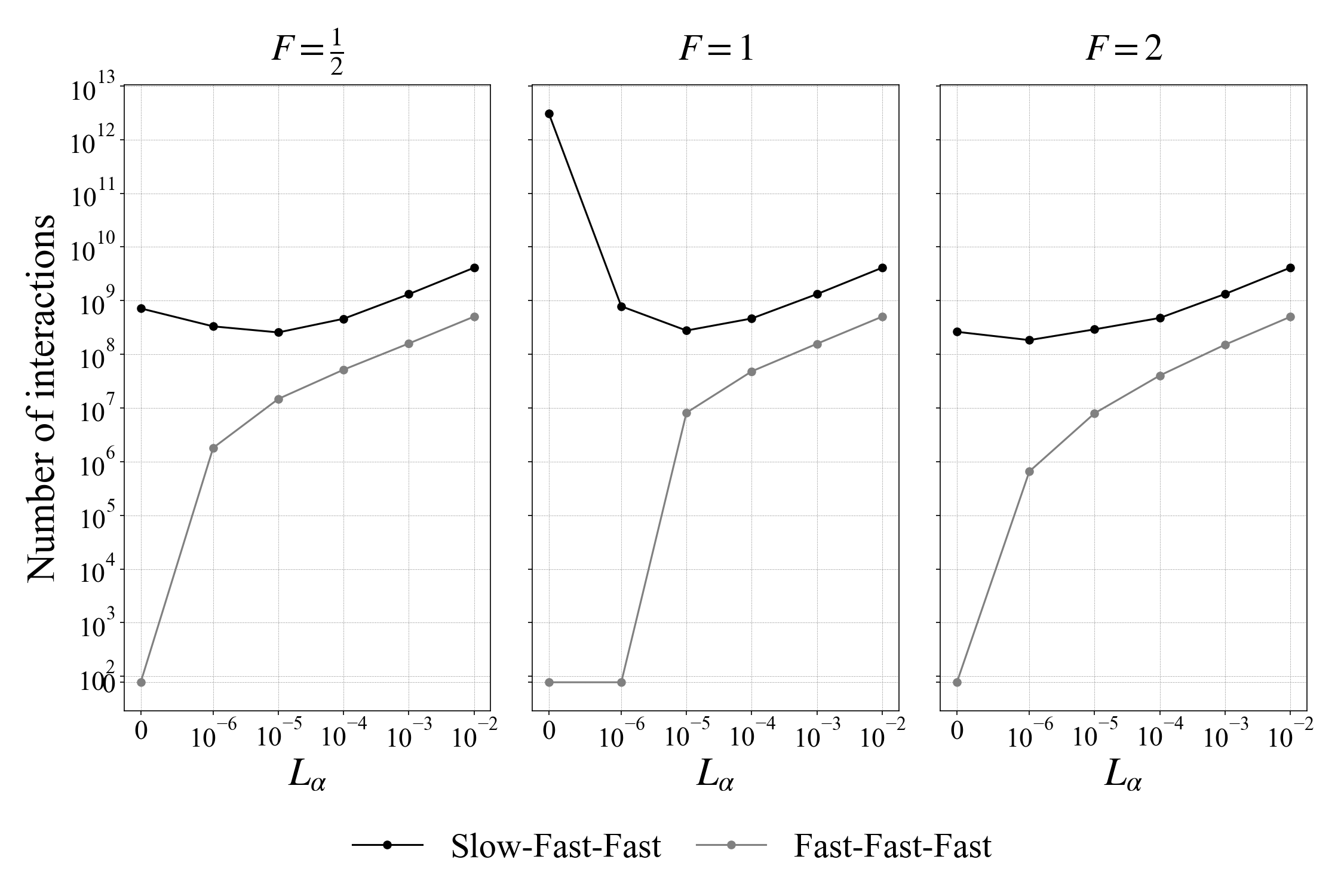}
    \caption{A detailed view of the change in the number of resonant interactions for small values of $L_\alpha$, for three values of the rotation-to-stratification ratio $F=f/N$ ($F=1/2,1,2$), shows that the number of fast-fast-fast triads appear to grow continuously with $L_\alpha$, rather than emerging in a discontinuous jump. Concurrently, the count of slow-fast-fast interactions exhibits an inflection point with $L_\alpha$. These counts are computed in the same manner as those in Figure \ref{fig:2}, but with respect to a range of smaller $L_\alpha$ values.}
    \label{fig:3}
\end{figure}
\par Fast-fast-fast exact resonant interactions do not occur when $1/2\leq F\leq 2$ in the $\alpha=0$ case, and resonant and near-resonant interactions occur only rarely otherwise. The condition is
\begin{equation}\label{eq:fast-fast-fast-cond-alphazero}
    \pm \frac{(|\bm{k}_H|^2+F^2k_3^2)^{\frac{1}{2}}}{|\bm{k}|} \pm   \frac{(|\bm{p}_H|^2+F^2p_3^2)^{\frac{1}{2}}}{|\bm{p}|}= \pm  \frac{(|\bm{q}_H|^2+F^2q_3^2)^{\frac{1}{2}}}{|\bm{q}|}.
\end{equation}
When $\alpha \neq 0$, \eqref{eq:fast-fast-fast-cond-alphazero} becomes instead
\begin{equation}\label{eq:fast-fast-fast-cond-alpha}
\begin{split}
    \pm \frac{((1+L_\alpha|\bm{k}|^2)|\bm{k}_H|^2+F^2 k_3^2)^{\frac{1}{2}}}{(1+L_\alpha|\bm{k}|^2)|\bm{k}|}    \pm &\frac{((1+L_\alpha|\bm{p}|^2)|\bm{p}_H|^2+F^2 p_3^2)^{\frac{1}{2}}}{(1+L_\alpha|\bm{p}|^2)|\bm{p}|} \\&=  \pm \frac{((1+L_\alpha|\bm{q}|^2)|\bm{q}_H|^2+F^2 q_3^2)^{\frac{1}{2}}}{(1+L_\alpha|\bm{q}|^2)|\bm{q}|}.
\end{split}
\end{equation}
\par The new nonlinearity and dependence on wavenumber in \eqref{eq:fast-fast-fast-cond-alpha} allows the number of fast-fast-fast resonances to significantly increase, even in cases where it is usually zero without $\alpha$-regularization. When $\alpha \neq 0$, the number of fast-fast-fast resonant interactions becomes comparable to the number of slow-fast-fast resonant interactions (Figure \ref{fig:2}). This stands in stark contrast to the usual fast QG limit, where fast-fast-fast triads are typically sparse or nonexistent. The emergence of a large number of fast-fast-fast resonances allows energy to be efficiently exchanged and redistributed purely among the fast modes, independent of the slow vortical flow. A detailed view of this emergence (Figure \ref{fig:3}) confirms that the number of fast-fast-fast triads appears to grow continuously from $L_\alpha=0$ rather than spiking discontinuously; this is especially evident in the $F=1/2$ ($Bu=4$) and $F=2$ ($Bu=1/4$) plots.

\par Perhaps most significantly, as $\alpha$ becomes sufficiently large, the numbers of both slow-fast-fast triads and fast-fast-fast triads converge to values largely independent of the rotation-to-stratification ratio $F$, indicating that $\alpha$ supersedes $F$ as the dominant parameter shaping the potential for resonant interactions. The overall effect of $\alpha$ on the fast dynamics is not to simply smooth but to restructure the energy pathways.
\section{Numerical simulations} 
\subsection{Description of the numerical method}
For all numerical experiments, we use Dedalus, a pseudo-spectral solver in Python \citep{burns2020}. The non-dimensional LANS-$\alpha$ equations \eqref{eq:lans-nondim-momentum}-\eqref{eq:lans-nondim-incompressibility} are solved within the triply periodic domain $[0,1]^3$, discretized on a uniform grid of $N^3=256^3$ points with a $3/2$ dealiasing rule. The system is initialized from a state of rest.  For time-stepping, we employ an explicit second-order, two-stage Runge-Kutta scheme (RK222). This scheme was selected for computational efficiency and validation runs confirmed its results were comparable to those from the third-order, four-stage scheme (RK443). The time step is adjusted dynamically by a CFL condition. 
\par The physical setup of our experiments is similar to that of \cite{smith2002}. Energy is injected into the system via a stochastic forcing that is constant in time. The forcing spectrum, $F(k)$, is a Gaussian in wavenumber space, centered at the forcing wavenumber $k_f$:
\begin{equation}
    F(k) = \frac{\eps_f}{(2\pi)^{\frac{1}{2}}s}\exp\left(\frac{-(k-k_f)^2}{2s^2}\right).
\end{equation}
The energy injection rate is $\eps_f=1$ and the standard deviation is $s=1$ for all simulations. Since we are interested in the quasi-geostrophic limit, we use low wavenumber forcing, $k_f=3(2\pi)$. This forcing defines a characteristic eddy turnover time, 
\begin{equation}
    \tau = (\eps_f k_f^2)^{-\frac{1}{3}}.
\end{equation}
All simulations are integrated to a final non-dimensional time of $t=1$, corresponding to approximately $7\tau$. This duration is sufficient for observing the initial development of the turbulent flow involving the transfer of energy between vortical and wave modes. In \eqref{eq:lans-nondim-momentum}, we have hyperviscosity with $n=8$, as in \cite{smith2002}, except the coefficient is constant in time and defined based on the dealiased grid cutoff wavenumber, in order to allow for more direct comparisons across simulations of different grid sizes. 

\par We investigate the system's behavior for three values of $F$: 1 (equally strong rotation and stratification), 2 (stronger rotation), and 1/2 (stronger stratification), achieved by setting $(Ro, Fr) = (0.1, 0.1)$, $(0.1, 0.2)$, and $(0.2, 0.1)$ respectively. In all cases, both $Ro$ and $Fr$ are small in order to reflect the QG limit. For each $F$, we systematically vary the nondimensional regularization parameter across five values: $L_\alpha = 0, 0.01, 0.1, 0.15, 0.2$. 

\subsection{Evolution of slow vortical and fast wave energy}
\begin{figure}[t]
    \centering
    \includegraphics[width=\linewidth]{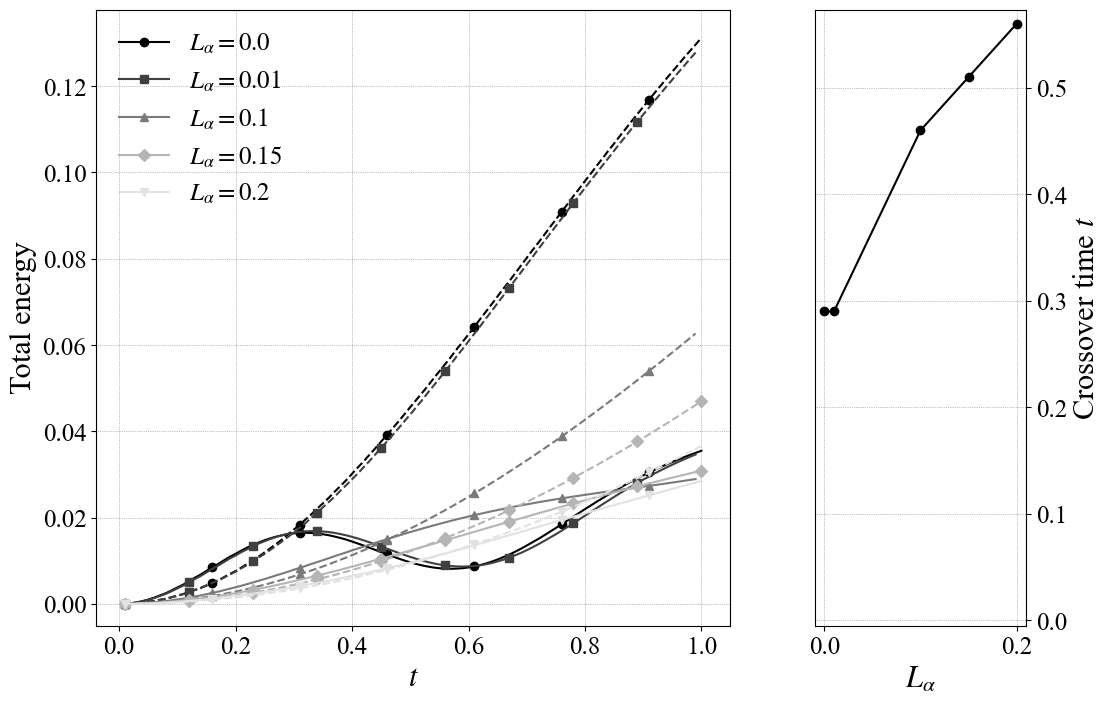}
    \caption{The regularization parameter $L_\alpha$ extends the lifetime of wave energy by delaying the dominance of vortical modes, as shown here for runs with $F=1$ ($Bu=1$). The left panel shows the decomposition of domain-integrated total energy (kinetic plus potential) into slow vortical energy (dashed) and wave energy (solid) for increasing $L_\alpha$. The right panel quantifies this delay, showing that the crossover time (when vortical energy surpasses wave energy) is a monotonically increasing function of $L_\alpha$.}
    \label{fig:4}
\end{figure}
\begin{figure}[!h]
    \centering
\includegraphics[width=\linewidth]{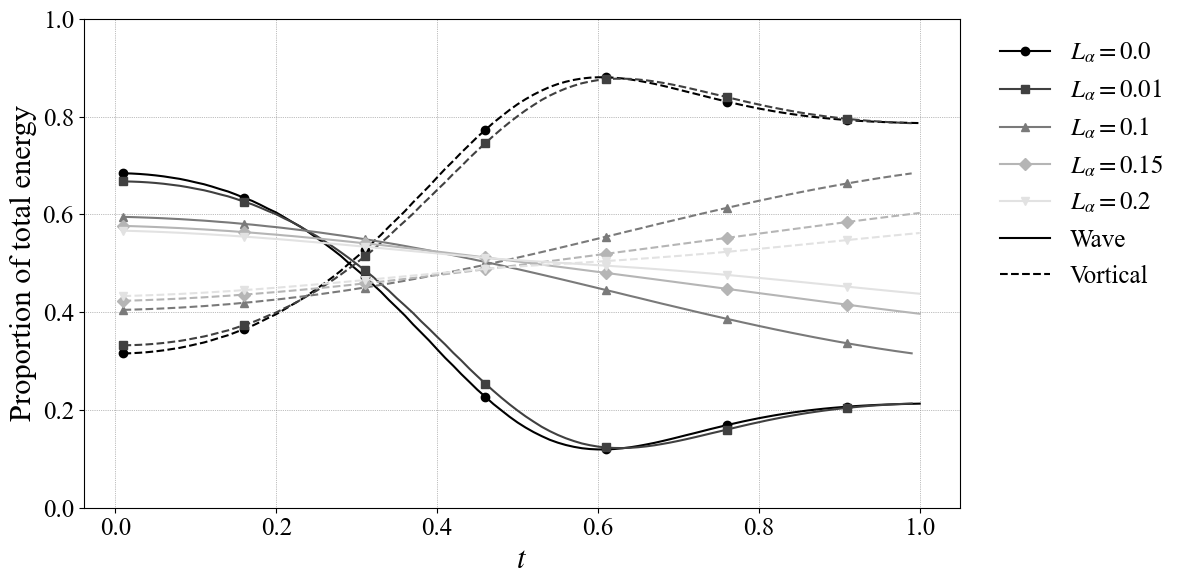}
    \caption{Energy decomposition for $F=1$ ($Bu=1$) showing the proportions of the domain-integrated total energy lying in slow vortical modes (dashed) and fast wave modes (solid) over time, corresponding to the magnitudes in Figure \ref{fig:4}. The crossover time is delayed with increasing $L_\alpha$.}
    \label{fig:5}
\end{figure}
The numerical simulations illustrate how the modifications to the set of resonant triads analyzed in $\S$\ref{subsec:fast-dynamics} manifest in the flow's energy evolution.  To see this, we track the partition of total energy between the vortical modes of the slow dynamics and the fast wave modes. Note that the total energy for the $L_\alpha \neq 0$ cases is given by $\frac{1}{2}\int_\Omega (\bm{u}\cdot \bm{v} + \rho^2) \,\text{d}\bm{x}$ as opposed to $\frac{1}{2}\int_\Omega (\bm{v}\cdot \bm{v} + \rho^2) \,\text{d}\bm{x}$, as the former is the conserved quantity in \eqref{eq:total-energy-conservation}. The former is smaller in magnitude than the latter, especially for large $L_\alpha$, because $\bm{u}$ is modified according to $\bm{v} = (1-L_\alpha^2\Delta)\bm{u}$ (the energy $\bm{v}\cdot\bm{v}$ for the $L_\alpha\neq 0$ cases better illustrate how the LANS-$\alpha$ simulations are indeed more energetic overall). 
\begin{figure}[t]
    \centering
    \includegraphics[width=\linewidth]{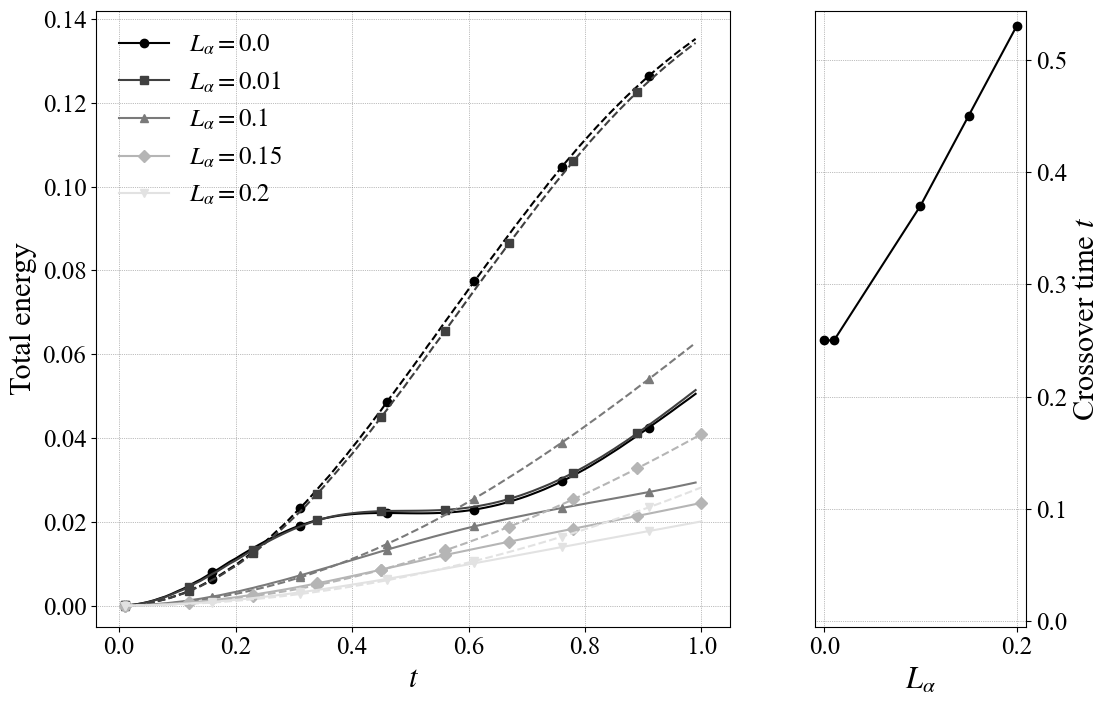}
    \caption{The decomposition of domain-integrated total (kinetic plus potential) energy into slow vortical energy (dashed) and fast wave energy (solid) for increasing $L_\alpha$, for $F=1/2$ ($Bu=4$). The right panel quantifies this delay, showing that the crossover time (when vortical energy surpasses wave energy) is a monotonically increasing function of $L_\alpha$.}
    \label{fig:6}
\end{figure}
\begin{figure}[!h]
    \centering
\includegraphics[width=\linewidth]{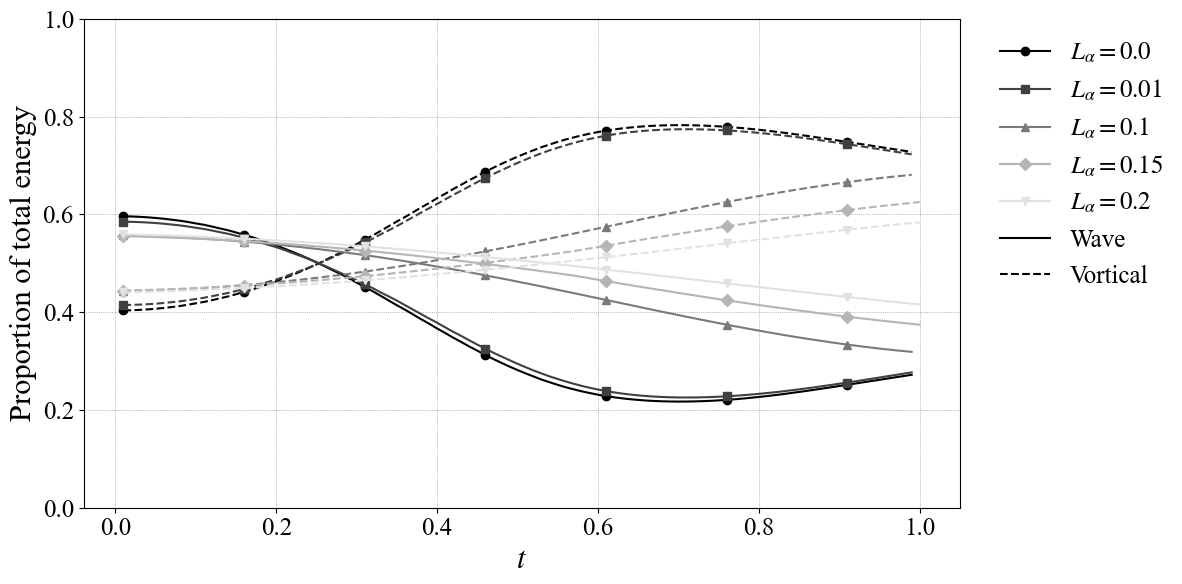}
    \caption{Energy decomposition for $F=1/2$ ($Bu=4$) showing the proportions of the domain-integrated total energy lying in slow vortical modes (dashed) and fast wave modes (solid) over time, corresponding to the magnitudes in Figure \ref{fig:6}.}
    \label{fig:7}
\end{figure}
\par The decomposition is computed by performing an eigenvalue decomposition of the fields outputted by the simulations, as in \cite{smith2002}, according to the eigenvalues in $\S$\ref{subsec:fast-dynamics}. This method is equivalent to applying the projection operator described in $\S$\ref{subsec:slow-dynamics}.

\par
\begin{figure}[t]
    \centering
    \includegraphics[width=\linewidth]{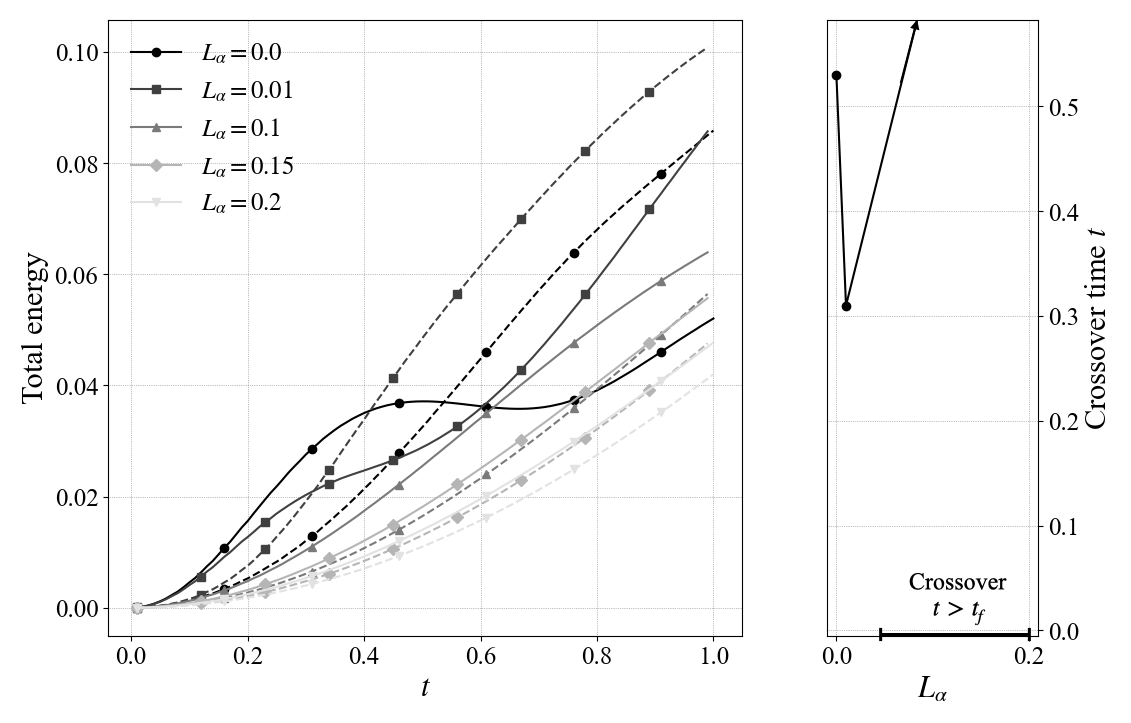}
    \caption{The decomposition of domain-integrated total (kinetic plus potential) energy into slow vortical energy (dashed) and fast wave energy (solid) for increasing $L_\alpha$, for $F=2$ ($Bu=1/4$). The right panel shows that in this case, the crossover time (when vortical energy surpasses wave energy) is no longer a monotonically increasing function of $L_\alpha$ and instead has an inflection point. When $L_\alpha$ exceeds 0.01, there is no crossover within the simulation period (non-dimensional time $t_f=1$), which is why the crossover times for those $L_\alpha$ values exceed the bounds of the plot. In all other simulations, the crossover time was achieved well within the simulation time, and there was no inflection point.}
    \label{fig:8}
\end{figure}

 The most immediate effect of the changes in the fast dynamics, as shown in the total energy plots for each value of $F$ (Figures \ref{fig:4}, \ref{fig:6}, and \ref{fig:8}), is a consistent delay in the evolution towards a state dominated by slow vortical modes. For all values of $F$ considered, increasing $L_\alpha$ systematically extends the crossover time at which the energy in vortical modes surpasses that in the wave modes (right-hand panels of Figures \ref{fig:4}, \ref{fig:6}, and \ref{fig:8}). This directly corresponds to a delay of geostrophic adjustment, and by hindering the flow's relaxation towards the balanced state, more energy is retained in the wave modes for a longer duration. 

\begin{figure}[h]
    \centering
\includegraphics[width=\linewidth]{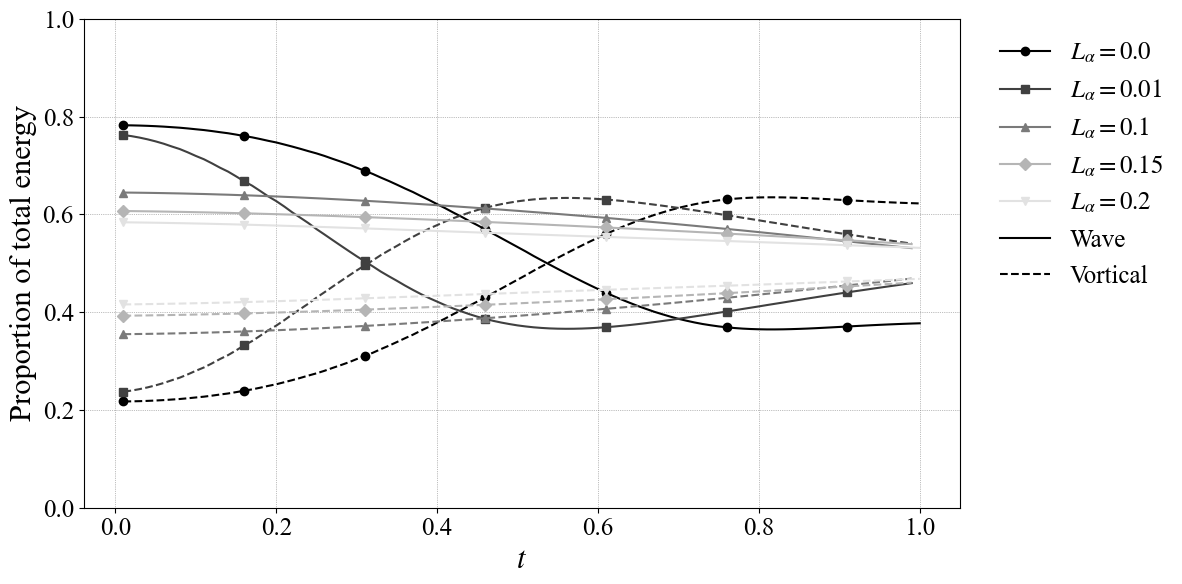}
    \caption{Energy decomposition for $F=2$ ($Bu=1/4$) showing the proportions of domain-integrated total energy lying in slow vortical modes (dashed) and fast wave modes (solid) over time, corresponding to the magnitudes in Figure \ref{fig:8}.}
    \label{fig:9}
\end{figure}
\par The emergence of fast-fast-fast resonant triads offers a potential mechanism to explain this delay. As shown in $\S$\ref{subsec:fast-dynamics}, the introduction of a nonzero $L_\alpha$ creates a dense network of these interactions, which are sparse or nonexistent in the $L_\alpha=0$ case. This new resonant pathway allows for the efficient exchange and redistribution of energy purely among the wave modes. Such a self-sustaining wave field, where energy is actively transferred among fast modes rather than being propagated away, is consistent with the observed delay in the system's relaxation to a geostrophic mode-dominated state. The creation of this fast-fast-fast pathway is a universal outcome of the regularization across all values of $F$, contrasting with the disruption of slow-fast-fast triads that is expected to be most impactful only in the $F=1$ ($Bu=1$) case. For the $Bu=1$ case, where geostrophic adjustment is extremely efficient in the absence of regularization, the disruption of this slow-fast-fast pathway likely plays an important, complementary role in producing the observed delay (Figures \ref{fig:4} and \ref{fig:5}). 
\par Although the regularization parameter $L_\alpha$ becomes the dominant factor (as opposed to $F$) shaping the fast dynamics (e.g., Figure \ref{fig:2}), the system's energy evolution still exhibits clear dependencies on the relative strengths of rotation and stratification. In the unregularized case, the dynamics vary considerably, as expected: the $F=2$ flow (Figure \ref{fig:9}) is most persistently wave-dominated, while the $F=1/2$ case (Figure \ref{fig:7}) becomes dominated by the slow modes most rapidly. Yet, the response to the regularization is also not uniform. For $F=1/2$ and $F=1$, the crossover time is a monotonically increasing function of $L_\alpha$ (Figures \ref{fig:6} and \ref{fig:4}). The $F=2$ case, however, exhibits a non-monotonic response, where a small amount of regularization ($L_\alpha=0.01$) actually accelerates the onset of vortical dominance before larger values delay it. 
\par The regularization parameter $L_\alpha$ acts as a second control on the system's energetics, in the sense that it governs the evolution of the partition of energy in a manner analogous to $F$ (and $Bu$). While higher values of both parameters ultimately limit the maximum proportion of vortical energy (see \cite{sukhatmesmith2008} for a comparison of the impact of many different values of $F$ in the $\alpha=0$ case), their effects on the initial distribution of wave and vortical energy are opposite. As shown in Figures \ref{fig:5}, \ref{fig:7}, and \ref{fig:9}, higher values of $L_\alpha$ lead to a smaller initial portion of wave energy -- the opposite effect from that of $F$. 
\par Over longer timescales, the growth in total energy is primarily driven by the slow vortical modes. This is evident from the steeper slope of the vortical energy curves after the initial adjustment period. However, the wave energy does not simply saturate or decay after the crossover. While in some cases it does appear to stagnate for brief periods, it generally continues to increase throughout the simulations, albeit at a slower rate than its vortical counterpart. 
\par These results show that $\alpha$ acts as a second control parameter for the system's energetic pathways, with effects that complement and sometimes oppose those of the rotation-to-stratification ratio, delaying geostrophic adjustment and sustaining energy growth in both wave and vortical components. 

\section{Conclusions}
\par In this paper, we derive the fast QG limit for the LANS-$\alpha$ model to analyze its energy evolution. In the absence of forcing and dissipation, the slow dynamics amount to the conservation of PV-$\alpha$. We find that $\alpha$ has the expected smoothing effect on these dominant, slow dynamics, but also that it amplifies the role of the fast wave dynamics. The regularization parameter $\alpha$ alters the frequencies of the waves and, in turn, the conditions for three-wave resonances, which play a key role in the distribution and transfer of energy in the system. By numerically testing integer wavenumbers, we find that introducing $\alpha$ allows for a rapid increase in the number of resonant triads where two fast waves produce a third fast wave. We also find that the number of slow-fast-fast resonant interactions greatly decreases in the $F=1$ case, unlike the creation of a large number of fast-fast-fast resonant interactions, which is universal across different values of $F$ (Figures \ref{fig:2} and \ref{fig:3}). Since the fast and slow dynamics are decoupled, we derive an energy conservation equation partitioned into slow and fast energies. Numerical simulations of the LANS-$\alpha$ system for three values of the rotation-to-stratification ratio $F$ and a range of $\alpha$ ($L_\alpha$) values produce fields that, once decomposed into slow and fast portions, exhibit energy dynamics consistent with this modification to the resonances.
\par Perhaps most significantly, as $\alpha$ becomes sufficiently large, the numbers of both slow-fast-fast triads and fast-fast-fast triads converge to values largely independent of $F$, indicating that $\alpha$ supersedes $F$ (and the Burger number) as the dominant parameter shaping the potential for resonant interactions. This reshaping both fundamentally alters the pathway for geostrophic adjustment and creates a rich wave-wave interaction regime. Indeed, the inflection point seen in the slow-fast-fast triad count (Figure \ref{fig:3}) 
(and in the crossover time in Figure \ref{fig:8}) may hint at a critical threshold for this behavior; exploring whether this value of $L_\alpha$ separates a stable regime from an unstable one, and its dependence on $F$, merits further investigation. 

Beyond its implications for the physics of the LANS-$\alpha$ model, this surge of fast wave interactions points to a potential physical origin for the model's numerical instabilities. We propose that such a dense field of interacting waves poses a challenge to numerical schemes because, for example, numerical discretizations can cause dispersion relations to produce both physical modes and computational modes, and the latter can lead to numerical instabilities \citep{dukowicz1994}. A model whose physics permits many more interactions between the physical modes may, in its numerical implementation, also provide more opportunities for the computational modes to interact with each other and the physical modes to trigger numerical instability. This supports the hypothesis that the origin for the known numerical instabilities of the LANS-$\alpha$ and related models is nonlinear, consistent with linear analyses indicating the model is stable (e.g., \citealt{hecht2008}). In fact, a (linear) stability analysis of the shallow water LANS-$\alpha$ model indicates that its numerical implementation should be \emph{more} stable than that of its unregularized counterpart in the sense that it permits a larger maximum allowable timestep as $\alpha$ increases \citep{wingate2004}. However, such linear analyses do not account for how numerical errors in individual waves can combine via triad interactions to further compromise the accuracy of the nonlinear dynamics, which can be a significant source of error depending on the choice of time-stepping scheme \citep{andrews2023}. Our results on enhanced wave-wave interactions indicate that future work, when using the LANS-$\alpha$ model or related models \citep{portamanazanna2014,anstey2017deformation}, could not only consider the inherent stability of the computational modes (e.g., the thorough analysis in \citealt{shchepetkin2005}), but also how they interact with other computational or physical modes in the context of fast-fast-fast interactions.

Ultimately, this analysis demonstrates that by reshaping the conditions for resonance, the LANS-$\alpha$ model permits a previously inaccessible regime of fast wave interactions which in turn affect the model's energy evolution, though the full implications of these dynamics and how they change in other relevant parameter regimes merit further study.

\backsection[Acknowledgements]{The authors thank Bruce Sutherland for valuable feedback on an earlier version of this manuscript.}

\backsection[Funding]{For the purpose of open access, the author has applied a ‘Creative Commons Attribution' (CC BY) license to any Author Accepted Manuscript version arising from this submission. This work was funded by the Geophysical Fluid Dynamics Fellowship at Woods Hole Oceanographic Institution. LRS was also supported by the Department of Defense (DoD) through the National Defense Science and Engineering Graduate (NDSEG) Fellowship Program. BW's research was also supported by the UK Engineering Physical Science Research Council (EPSRC) grant number EP/R029628/1, Leverhulme Trust Research Fellowship RF-2022-013.}

\backsection[Declaration of interests]{The authors report no conflict of interest.}

\backsection[Data availability statement]{The code used to produce all simulations and analysis for this study is available on GitHub at \url{https://github.com/lulabels/lans-alpha} and has been archived at \url{https://doi.org/10.5281/zenodo.17188052}.}

\backsection[Author ORCIDs]{L. R. Seitz, \url{https://orcid.org/0009-0004-1933-1931}, Beth A. Wingate \url{https://orcid.org/0000-0003-2464-6132}.}

\begin{appen}

\section{Derivation of the slow dynamics}\label{appA}
\subsection{Derivation of the projection operator}\label{app:deriv-proj-op}
In this section, we derive the formula for the projection onto the null space of the fast operator as in \cite{whiteheadwingate2014}. In the QG limit, the fast operator is given by the linear combination $\mathcal{L}_{Fr}+F\mathcal{L}_{Ro}$, with these component operators defined in \eqref{eq:lans-abstract-operators}. We take Fourier transform and find the the matrix for this fast operator:
\begin{equation}\label{eq:QG-fast-matrix}
A(\bm{k}) = \frac{1}{|\bm{k}|^2} \begin{bmatrix} -\frac{k_1k_2}{Ro(1+L_\alpha|\bm{k}|^2)} & -\frac{k_2^2+k_3^2}{Ro(1+L_\alpha|\bm{k}|^2)} & 0 & -\frac{k_1k_3}{Fr} \\ \frac{k_1^2+k_3^2}{Ro C} & \frac{k_1k_2}{Ro(1+L_\alpha|\bm{k}|^2)} & 0 & -\frac{k_2k_3}{Fr} \\ -\frac{k_2k_3}{Ro(1+L_\alpha|\bm{k}|^2)} & \frac{k_1k_3}{Ro(1+L_\alpha|\bm{k}|^2)} & 0 & \frac{k_1^2+k_2^2}{Fr} \\ 0 & 0 & -\frac{|\bm{k}|^2}{Fr(1+L_\alpha|\bm{k}|^2)} & 0 \end{bmatrix}.
\end{equation}
The null space is spanned by the unit vector
\begin{equation}\label{eq:QG-fast-null-unit-vec}
\left({k_1^2+k_2^2+\left(\frac{F}{1+L_\alpha|\bm{k}|^2}\right) ^2 k_3^2}\right)^{-\frac{1}{2}}\begin{bmatrix} k_2 \\ -k_1 \\ 0 \\ \frac{F}{1+L_\alpha|\bm{k}|^2} k_3\end{bmatrix}.
\end{equation}
Then we can construct, where $\hat{\bm{w}}_k=(\hat{u}_k, \hat{v}_k, \hat{w}_k, \hat{\rho}_k)$,
\begin{equation}\label{eq:QG-fast-proj-vec}
P\hat{\bm{w}}_k = \frac{1}{k_1^2+k_2^2+\left(\frac{F}{1+L_\alpha|\bm{k}|^2}\right)^2k_3^2}\begin{bmatrix}k_2^2\hat{u}_k-k_1k_2\hat{v}_k+\frac{F}{1+L_\alpha|\bm{k}|^2}k_2k_3\hat{\rho}_k  \\ k_1^2\hat{v}_k-k_1k_2\hat{u}_k-\frac{F}{1+L_\alpha|\bm{k}|^2}k_1k_3\hat{\rho}_k  \\ 0 \\ \frac{F}{1+L_\alpha|\bm{k}|^2} k_3\left(k_2\hat{u}_k-k_1\hat{v}_k+\frac{F}{1+L_\alpha|\bm{k}|^2}k_3\hat{\rho}_k\right) \end{bmatrix}.
\end{equation}
To transform back into physical space, one simply adds and subtracts $(k_1^2+F^2(1+L_\alpha|\bm{k}|^2)^{-2}k_3^2)\hat{u}_k$ from the first component of \eqref{eq:QG-fast-proj-vec} and $(k_2^2+F^2(1+L_\alpha|\bm{k}|^2)^{-2}k_3^2)\hat{v}_k$ from the second. This produces a cancellation with the denominator that yields the $\bm{v}_H$ term in \eqref{qg-proj-alpha-neq0} (notice this does not have any $\mathcal{S}^{-1}$). Transforming all terms back into physical space yields \eqref{qg-proj-alpha-neq0}. 

\subsection{Fast-slow decomposition}\label{app:slow-fast-decomp}
In this section, we present additional details of the fast-slow decomposition \eqref{eq:slow-fast-decomp} that will inform how to use the projection to derive the slow dynamics (completed in Appendix \ref{app:qg-pv-alpha-deriv}). In \eqref{eq:abstract-operator-form}, we can identify the fast operator as $\mathcal{L}_F  = F\mathcal{L}_{
Ro} + \mathcal{L}_{Fr}$, which is multiplied by $\frac{1}{\eps}$, where $\eps$ is the Rossby number. The solution to \eqref{eq:abstract-operator-form}, which is based on the method of multiple scales, will be denoted $\bm{w}^\eps$ and depends on both the fast time scale $\tau = \frac{t}{\eps}$ and the slow time scale $t$. The solution can be expanded as 
\begin{equation}\label{eq:solution-decomposition}
    \bm{w}^\eps(\bm{x},t,\tau) = \bm{w}^0(\bm{x},t,\tau) + \eps \bm{w}^1(\bm{x},t,\tau). 
\end{equation}
 Substituting \eqref{eq:solution-decomposition} into the abstract operator form \eqref{eq:abstract-operator-form}, we obtain the $O(\eps^{-1})$ equation 
\begin{equation}\label{eq:abstract-op-oepsminus1}
    \frac{\partial \bm{w}^0}{\partial \tau} + \mathcal{L}_F(\bm{w}^0)=0.
\end{equation}
By separation of variables, \eqref{eq:abstract-op-oepsminus1} has a solution of the form
\begin{equation}\label{eq:abstract-op-oepsminus1-sol}
\bm{w}^0(\bm{x},t,\tau)=e^{-\tau \mathcal{L}_F}\overline{\bm{w}}(\bm{x},t).
\end{equation}
At next order, we obtain the $O(\eps^0)$ equation 
\begin{equation}\label{eq:abstract-op-oepzero}
    \frac{\partial \bm{w}^1}{\partial \tau}+\mathcal{L}_F(\bm{w}^1)
 = -\left(\frac{\partial \bm{w}^0}{\partial \tau} + \mathcal{L}_S(\bm{w}^0)+\mathcal{B}(\bm{w}^0,\bm{w}^0)-\mathcal{D}(\bm{w}^0)\right).\end{equation}
 Using Duhamel's principle and \eqref{eq:abstract-op-oepsminus1-sol}, \eqref{eq:abstract-op-oepzero} has a solution of the form 
 \begin{equation}\label{eq:fast-sol-with-int}
 \begin{split}
     e^{\tau \mathcal{L}_F}\bm{w}^1 = \bm{w}^1\Big|_{\tau = 0} - \tau\frac{\partial \overline{\bm{w}}}{\partial t}(\bm{x},t)
- \int_0^\tau e^{s\mathcal{L}_F}&(\mathcal{L}_S(e^{-s\mathcal{L}_F}\overline{\bm{w}})+\bilform{e^{-s\mathcal{L}_F}\overline{\bm{w}}}{e^{-s\mathcal{L}_F}\overline{\bm{w}}} - \mathcal{D}(e^{-s\mathcal{L}_F}\overline{\bm{w}}))\,\text{d}s.
\end{split}
\end{equation}
The integral in \eqref{eq:fast-sol-with-int} is zero due to \cite{schochet1994}. Then due to the identification \eqref{eq:abstract-op-oepsminus1-sol}, the principal term $\overline{w}(\bm{x},t)$ has no fast oscillations. Overall, to order $\eps$, 
\begin{equation}\label{eq:solution-decomposition-2}
\bm{w}^\eps(\bm{x},t) = e^{-\frac{t}{\eps}\mathcal{L}_F}\overline{\bm{w}}(\bm{x},t) + o(1),
 \end{equation}
 valid as $\eps \to 0$. While $\overline{\bm{w}}$ has no oscillations, $e^{-\frac{t}{\eps}\mathcal{L}_F}\overline{\bm{w}}(\bm{x},t)$ does. This motivates the fast-slow decomposition. Since $\mathcal{L}_F$ is required to be skew-Hermitian, by the Spectral Theorem, we can decompose $\overline{\bm{w}}$ into a portion outside of the null space of $\mathcal{L}_F$ and a portion lying within the null space of $\mathcal{L}_F$. This decomposition can be interpreted as 
 \begin{equation}\label{eq:slow-fast-decomp-app}
     \overline{\bm{w}}(\bm{x},t) = \overline{\bm{w}}_F(\bm{x}, t) + \overline{\bm{w}}_S(\bm{x},t),
 \end{equation}
 where $\overline{\bm{w}}_F\in \text{Range }\mathcal{L}_F$ is the fast portion and $\overline{\bm{w}}_S\in\text{Kernel }\mathcal{L}_F$ is the slow portion. We have thus derived \eqref{eq:slow-fast-decomp} in detail. Substituting \eqref{eq:slow-fast-decomp-app} into \eqref{eq:solution-decomposition-2}, 
 \begin{equation}\label{eq:solution-decomposition-3}
\bm{w}^\eps(\bm{x},t) = e^{-\frac{t}{\eps}\mathcal{L}_F}\overline{\bm{w}}_F(\bm{x},t) + \overline{\bm{w}}_S(\bm{x},t)+o(1). 
 \end{equation}
 Then $\overline{\bm{w}}_S$ has no fast oscillations. Now we can find equations whose solutions are $\overline{\bm{w}}_F$ and $\overline{\bm{w}}_S$, respectively. We can find an equation for the slow operator by projecting \eqref{eq:abstract-operator-form} onto the null space of the fast operator. However, we can simplify this by projecting only the $O(\eps^0)$ version of \eqref{eq:abstract-operator-form}, which we have found so far to be 
\begin{equation}\label{eq:abstract-operator-form-o1}
    \frac{\partial \bm{w}^0}{\partial t} + \mathcal{L}_F(\bm{w}^1)+\mathcal{L}_S(\bm{w}^0)+\bilform{\bm{w}^0}{\bm{w}^0}-\mathcal{D}(\bm{w}^0)=0.
\end{equation} 
 Letting $P$ denote such a projection operator, then $\bm{w}_S$ may be alternatively represented as the solution to (\emph{i.e}. identified with $\bm{w}^0$ in)
 \begin{equation}\label{eq:slow-dynamics}
    \frac{\partial \bm{w}^0}{\partial t} + P(\mathcal{L}_S(\bm{w}^0)+\bilform{\bm{w}^0}{\bm{w}^0}-\mathcal{D}(\bm{w}^0))=0.
\end{equation} 

\subsection{Derivation of the conservation of three-dimensional QG PV-\texorpdfstring{$\alpha$}{alpha}}\label{app:qg-pv-alpha-deriv}
We use the projection derived in Appendix \ref{app:deriv-proj-op} to derive the slow dynamic equations, in the absence of viscosity. Before doing so, we establish a necessary vorticity identity. From \cite{holmfluctuations}, the vorticity equation for the LANS$-\alpha$ equations is
\begin{equation}\label{eq:lans-alpha-vort-holm}
\frac{\partial \bm{\omega}}{\partial t} + \mathcal{S}^{-1}\bm{v}\cdot \nabla \bm{\omega} = \bm{\omega} \cdot \nabla \mathcal{S}^{-1}\bm{v}.
\end{equation}
Here, the notation $\mathcal{S}^{-1}\bm{v}\cdot \nabla \boldsymbol{\omega}$ means multiplying a row vector $\mathcal{S}^{-1}\bm{v}$ with the gradient (matrix) of $\omega$. We require an analogous identity for the horizontal vorticity, i.e., the curl of the total advective operator (including the $\bm{v}\cdot \nabla \bm{u}^T$ term) but in two dimensions. An explicit calculation yields that, upon considering incompressibility,
\begin{equation}\label{lans-curl-identity}
    \nabla_H \times((\nabla_H \bm{v}_H)\bm{u}_H+\bm{v}_H(\nabla_H \bm{u}_H)) = (\mathcal{S}^{-1}\bm{v}_H \cdot \nabla_H)\omega_3.
\end{equation}
Reducing the projected equations to the conservation of potential vorticity crucially requires the application of hydrostatic and geostrophic balance, which can be seen to be formally valid even in the case of a fast singular limit (e.g., \citealt{majda-textbook}). The key equation to see that this generalizes in the $\alpha \neq 0$ case is the vorticity identity \eqref{lans-curl-identity}. From this, a generalized Ertel's potential vorticity theorem can be proven, which is the theorem required to formally show leading order hydrostatic and geostrophic balance.   
\par Now, to compute the slow dynamics equations from the projection, we use \eqref{eq:slow-dynamics}. Since the slow operator is zero in this limit and we are finding the slow dynamics in the absence of diffusion, we compute $P_{\text{QG}_\alpha}\mathcal{B}(\bm{w}, \bm{w})$, where, as before, $\bm{w} = (\bm{v}_H, w, \rho)^T$. We need to substitute in for $\mathcal{B}(\bm{v}_H, \bm{v}_H)$ (where this is the appropriately truncated operator):
\begin{equation}\label{eq:eval-inside-proj-1}
(\mathcal{S}^{-1}\bm{v}_H\cdot\nabla_H)\bm{v}_H-\nabla_H \Delta_H^{-1}\Big(\nabla_H \cdot ((\mathcal{S}^{-1}\bm{v}_H\cdot \nabla_H) \bm{v}_H)\Big).
\end{equation}
 Here, $\mathcal{S}^{-1}$ indicates the two-dimensional version of the operator, and there is an $H$ on the inverse Laplacian because it is actually a vector Laplacian. For $\mathcal{B}(\rho, \rho)$ (this is only the fourth component of the operator), 
\begin{equation}
(\mathcal{S}^{-1}\bm{v}_H\cdot \nabla_H)\rho.
\end{equation}
\par Upon applying the projection, the contribution from the second term in \eqref{eq:eval-inside-proj-1} will be zero. Then
\begin{align}
P_{\text{QG}_\alpha}\mathcal{B}(\bm{v}, \bm{v}) &=\Delta_{\text{QG}_\alpha}^{-1}\left(\nabla_H^\perp( \nabla_H \times ((\mathcal{S}^{-1}\bm{v}_H\cdot \nabla_H)\bm{v}_H))-F\mathcal{S}^{-1}\nabla_H^\perp\left(\frac{\partial}{\partial z}( (\mathcal{S}^{-1}\bm{v}_H\cdot \nabla_H)\rho)\right)\right) \nonumber \\
&= \Delta_{\text{QG}_\alpha}^{-1}\left(\nabla_H^\perp( (\mathcal{S}^{-1}\bm{v}_H\cdot \nabla_H)\omega_3) -F\mathcal{S}^{-1}\nabla_H^\perp\left(\frac{\partial}{\partial z}( (\mathcal{S}^{-1}\bm{v}_H\cdot \nabla_H)\rho)\right)\right) .
\end{align}
Using the commutativity of $\mathcal{S}^{-1}$ with the advection operator (under the interpretation of Lagrangian averaging, i.e., that this is a valid approximation to the dynamic Helmholtz operator), from which it follows due to the lack of time-dependence that there is commutativity with just the $\mathcal{S}^{-1}\bm{v}_H\cdot \nabla_H$ portion, we obtain 
\begin{equation}
P_{\text{QG}_\alpha}\mathcal{B}(\bm{v}, \bm{v})= \Delta_{\text{QG}_\alpha}^{-1}\left(\nabla_H^\perp( (\mathcal{S}^{-1}\bm{v}_H\cdot \nabla_H)\omega_3) -F\nabla_H^\perp\left(\frac{\partial}{\partial z} (\mathcal{S}^{-1}\bm{v}_H\cdot \nabla_H)\mathcal{S}^{-1}\rho)\right)\right). 
\end{equation}
We apply hydrostatic balance, which in this case yields that $\omega_3 = \mathcal{S}\Delta_H\phi$ and $\rho = -F\frac{\partial \phi}{\partial z}$. We can commute the partial derivative with respect to $z$ with the advective term due to the relationships with $\phi$. We obtain 
\begin{equation}
P_{\text{QG}_\alpha}\mathcal{B}(\bm{v}, \bm{v}) = \Delta_{\text{QG}_\alpha}^{-1}\left(\nabla_H^\perp( (\mathcal{S}^{-1}\bm{v}_H\cdot \nabla_H)\mathcal{S}\Delta_H\phi) +F^2\nabla_H^\perp\left( (\mathcal{S}^{-1}\bm{v}_H\cdot \nabla_H)\mathcal{S}^{-1}\frac{\partial^2}{\partial z^2}\phi\right)\right). 
\end{equation}
Now assembling the first row (velocity portion) of the slow equation \eqref{eq:slow-dynamics}, we have 
\begin{equation}
\frac{\partial\bm{v}_H}{\partial t} + \Delta_{\text{QG}_\alpha}^{-1}\left(\nabla_H^\perp( (\mathcal{S}^{-1}\bm{v}_H\cdot \nabla_H)\mathcal{S}\Delta_H\phi) +F^2\nabla_H^\perp\left( (\mathcal{S}^{-1}\bm{v}_H\cdot \nabla_H)\mathcal{S}^{-1}\frac{\partial^2}{\partial z^2}\phi\right)\right) =0.
\end{equation}
After applying $\Delta_{\text{QG}_\alpha}$ to both sides, and substituting in $\bm{v}_H = \mathcal{S} \nabla_H^\perp \phi$, we obtain 
\begin{equation}
\frac{\partial}{\partial t}\left(\Delta_H\nabla_H^\perp \mathcal{S} \phi + F^2 \mathcal{S}^{-2}\frac{\partial^2}{\partial z^2}\mathcal{S}\nabla_H^\perp\phi\right)+ \nabla_H^\perp( (\nabla_H^\perp \phi\cdot \nabla_H)\mathcal{S}\Delta_H\phi) +F^2\nabla_H^\perp\left( (\nabla_H^\perp \phi\cdot \nabla_H)\mathcal{S}^{-1}\frac{\partial^2}{\partial z^2}\phi\right) =0.
\end{equation}
After pulling out the $\nabla_H^\perp$ and either taking the horizontal curl of both sides in order to get a horizontal Laplacian and then applying the inverse horizontal Laplacian to both sides, or integrating each row and then putting together the results to see that the constant of integration must be zero, we obtain exactly \eqref{eq:qg-alpha-v1}.
\section{Derivation of the fast dynamics}\label{appB}
\subsection{Eigenvectors}
In this section, we write the formulas for the eigenvectors corresponding to \eqref{eq:eigenvalues-QG-alpha}, for the various cases of the wavenumber $\bm{k}$. Both the eigenvalues and eigenvectors are similar to those in \cite{smith2002} and the eigendecomposition that can be used to decompose the dynamics into fast and slow portions (after appropriately orthonormalizing the basis) is given by eq. (2.16) therein.
\par The most general case is when $\bm{k}\neq 0, \bm{k}_H\neq 0$:
\begin{subequations}
\begin{align}
    \bm{r}^{(1)}(\bm{k}) &= \frac{1}{\sqrt{2}|\bm{k}_H||\bm{k}|(1+L_\alpha|\bm{k}|^2)}\begin{bmatrix} i(1+L_\alpha|\bm{k}|^2)k_1k_3 - \frac{Fk_2k_3}{\omega^{(1)}(\bm{k})} \\ i(1+L_\alpha|\bm{k}|^2)k_2k_3 + \frac{Fk_1k_3}{\omega^{(1)}(\bm{k})} \\ -i(1+L_\alpha|\bm{k}|^2)|\bm{k}_H|^2 \\ \frac{|\bm{k}_H|^2} {\omega^{(1)}(\bm{k})} \end{bmatrix}, \\
\bm{r}^{(-1)}(\bm{k}) &= \frac{1}{\sqrt{2}|\bm{k}_H||\bm{k}|(1+L_\alpha|\bm{k}|^2)}\begin{bmatrix} -i(1+L_\alpha|\bm{k}|^2)k_1k_3 - \frac{Fk_2k_3}{\omega^{(1)}({\bm{k}})} \\ -i(1+L_\alpha|\bm{k}|^2)k_2k_3+ \frac{Fk_1k_3}{\omega^{(1)}(\bm{k})} \\ i(1+L_\alpha|\bm{k}|^2)|\bm{k}_H|^2 \\ \frac{|\bm{k}_H|^2} {\omega^{(1)}(\bm{k})} \end{bmatrix}, \\
\bm{r}^{(0)}(\bm{k}) &= \frac{1}{|\bm{k}|}\begin{bmatrix}-\frac{i(1+L_\alpha|\bm{k}|^2)k_2}{\omega^{(1)}(\bm{k})} \\ \frac{i(1+L_\alpha|\bm{k}|^2)k_1}{\omega^{(1)}(\bm{k})} \\ 0 \\ -\frac{iFk_3}{\omega^{(1)}(\bm{k})}\end{bmatrix}. 
\end{align}
\end{subequations}
Even though $\omega_{\bm{k}}^{(0)}$ is a double eigenvalue, there are only three eigenvectors because the last does not respect incompressibility. 
\par The second case is when $\bm{k}_H=0$ but $\bm{k}\neq 0$
\begin{equation}
 \bm{r}_{\bm{k}}^{(1)} = \begin{bmatrix} \frac{i}{\sqrt{2}} \\ \frac{1}{\sqrt{2}} \\ 0 \\ 0 \end{bmatrix}, \quad \bm{r}_{\bm{k}}^{(-1)} = \begin{bmatrix} -\frac{i}{\sqrt{2}} \\ \frac{1}{\sqrt{2}} \\ 0 \\ 0 \end{bmatrix}, \quad \bm{r}_{\bm{k}}^{(0)} = \begin{bmatrix} 0 \\ 0\\ 0 \\ 1 \end{bmatrix}.
 \end{equation}
The last case is when $\bm{k}=0$. In this case the four eigenfunctions all correspond to wave modes, and have frequencies $\omega^{(\pm 1)}(\bm{0}) = \pm 1, \tilde{\omega}^{(\pm 1)}(\bm{0}) = \pm F$. The eigenfunctions are 
\begin{equation}
 \bm{r}_{\bm{0}}^{(1)} = \begin{bmatrix} \frac{i}{\sqrt{2}} \\ \frac{1}{\sqrt{2}} \\ 0 \\ 0 \end{bmatrix}, \quad \bm{r}_{\bm{0}}^{(-1)} = \begin{bmatrix} -\frac{i}{\sqrt{2}} \\ \frac{1}{\sqrt{2}} \\ 0 \\ 0 \end{bmatrix}, \quad \tilde{\bm{r}}_{\bm{0}}^{(1)} = \begin{bmatrix} 0 \\ 0\\ \frac{1}{\sqrt{2}}\\ \frac{i}{\sqrt{2}} \end{bmatrix}, \quad \tilde{\bm{r}}_{\bm{0}}^{(-1)} = \begin{bmatrix} 0 \\ 0\\ \frac{1}{\sqrt{2}}\\ -\frac{i}{\sqrt{2}} \end{bmatrix}.
 \end{equation}
\subsection{Range of purely fast interactions}\label{appB2}
In $\S$\ref{subsec:fast-dynamics}, Table \ref{tab:resonances}, a criterion is given for when there are no fast-fast-fast interactions. When $\alpha=0$, there are none when $\frac{1}{2}\leq F\leq 2$ (c.f. \cite{smith2002}). Considering the nonzero eigenvalues \eqref{eq:eigenvalues-QG-alpha}, if $k_3=0$, then 
\begin{equation}
    \omega^{(\pm 1)}(\bm{k})=\pm (1+L_\alpha|\bm{k}_H|^2)^{-\frac{1}{2}}.
\end{equation}
If instead $|\bm{k}_H|^2=0$ then 
\begin{equation}
    \omega^{(\pm 1)}(\bm{k})=\pm \frac{F}{1+L_\alpha k_3^2}.
\end{equation}
Accordingly,  
\begin{equation}
    \min |\omega(\bm{k})| = \min\left((1+L_\alpha|\bm{k}_H|^2)^{-\frac{1}{2}}, \frac{F}{1+L_\alpha k_3^2}\right) \text{ and } \max |\omega(\bm{k})| = \max\left((1+L_\alpha|\bm{k}_H|^2)^{-\frac{1}{2}}, \frac{F}{1+L_\alpha k_3^2}\right). 
\end{equation}
For $\omega(\bm{k}) + \omega (\bm{p}) = \omega(\bm{q})$ (without loss of generality, $\omega(\bm{k}), \omega(\bm{p}), \omega(\bm{q})\geq 0$), it must be the case that $2\min |\omega | < \max |\omega |$. Thus, whenever
\begin{equation}
\frac{1}{2} \leq \frac{1+L_\alpha k_3^2}{F(1+L_\alpha|\bm{k}_H|^2)^{\frac{1}{2}}}\leq 2,
\end{equation}
there are no resonant triad interactions purely among inertial-gravity waves. 
\section{Derivation of energy conservation}\label{appC}
The quadratic invariants of the LANS-$\alpha$ equations appeared elsewhere (e.g., \citealt{foias2001}), but for completeness we provide the derivation of \eqref{eq:total-energy-conservation} here. To derive the energy conservation law \eqref{eq:total-energy-conservation}, we dot \eqref{eq:lans-nondim-momentum} with $\bm{u}$ and multiply \eqref{eq:lans-nondim-buoyancy} by $\rho$ (with $n=1$ in the viscous operator). We obtain the kinetic energy equation, which we integrate over the domain,
\begin{equation}\label{eq:energy-cons-deriv-1}
    \int_\Omega \bm{u} \cdot \frac{\partial\bm{v}}{\partial t} \,\text{d}\bm{x} + \int_\Omega \bm{u}\cdot(\bm{u}\cdot\nabla \bm{v} + \bm{v}\cdot\nabla \bm{u}^T) \,\text{d}\bm{x} + \int_\Omega \nabla \cdot(\phi\bm{u}) + \frac{1}{Fr}\rho\hat{\bm{z}}\cdot\bm{u} \,\text{d}\bm{x}  = \frac{1}{Re}\int_\Omega \bm{u}\cdot\Delta \bm{v} \,\text{d}\bm{x}. 
\end{equation}
Note that the rotation term vanished upon taking the dot product with $\bm{u}$. For the potential energy equation, we obtain 
\begin{equation}\label{eq:energy-cons-deriv-pe-1}
    \int_\Omega \rho \frac{\partial \rho}{\partial t} \,\text{d}\bm{x}+ \int_\Omega \rho(\bm{u}\cdot \nabla \rho) \,\text{d}\bm{x}- \int_\Omega \frac{1}{Fr} \rho (\bm{u}\cdot \hat{\bm{z}})\,\text{d}\bm{x} = \frac{1}{RePr}\int_\Omega \rho \Delta \rho \,\text{d}\bm{x}.
\end{equation}

\par We are considering triply periodic functions (which are sufficiently regular, on a sufficiently regular, bounded domain). Consequently, the inverse Helmholtz operator is self-adjoint because the Helmholtz operator is, which is easily seen via the application of integration by parts twice. Using this property,  
\begin{equation}\label{eq:energy-cons-time}
    \frac{d}{dt}\int_\Omega \frac{1}{2}(\bm{u}\cdot\bm{v})\,\text{d}\bm{x} = \frac{1}{2}\int_\Omega \bm{u}\cdot \frac{\partial \bm{v}}{\partial t} + \bm{v}\cdot \frac{\partial \bm{u}}{\partial t} \,\text{d}\bm{x}= \int_\Omega \bm{u}\cdot \frac{\partial \bm{v}}{\partial t} \,\text{d}\bm{x}.
\end{equation}
To evaluate the advective term, we write it in index notation:
\begin{equation}
    \int_\Omega \bm{u}\cdot(\bm{u}\cdot\nabla \bm{v} + \bm{v}\cdot\nabla \bm{u}^T) \,\text{d}\bm{x} = \int_\Omega u_i(u_j\partial_jv_i+v_j\partial_iu_j) \,\text{d}\bm{x}= \int_\Omega u_iv_j\partial_iu_j - v_i((\partial_ju_i)u_j+u_i(\partial_ju_j)) \,\text{d}\bm{x}.
\end{equation}
Hence 
\begin{equation}\label{eq:energy-cons-advec-1}
  \int_\Omega \bm{u}\cdot(\bm{u}\cdot\nabla \bm{v} + \bm{v}\cdot\nabla \bm{u}^T)  =\int_\Omega u_iv_j\partial_iu_j -\int_\Omega v_iu_j\partial_ju_i =0. 
\end{equation}
Here we used the divergence theorem twice and evaluated the difference to be zero by relabeling the dummy indices. The pressure term in \eqref{eq:energy-cons-deriv-1} is zero due to the divergence theorem. For the advection term in \eqref{eq:energy-cons-deriv-pe-1}, 
\begin{equation}\label{eq:energy-cons-advec-2}
    \int_\Omega \rho(\bm{u}\cdot \nabla \rho) \,\text{d}\bm{x}= \int_\Omega \bm{u}\cdot\nabla \left(\frac{1}{2}\rho^2\right)  \,\text{d}\bm{x}= \int_\Omega\nabla \cdot \left(\bm{u}\cdot \frac{1}{2}\rho
    ^2\right)  \,\text{d}\bm{x}- \int_\Omega(\nabla \cdot \bm{u})\left(\frac{1}{2}\rho^2\right) \,\text{d}\bm{x}=0,
\end{equation}
where we used the product rule to split the integral and divergence theorem and incompressibility to conclude the penultimate expression equals zero. 
For the diffusive term, 
\begin{equation}\label{eq:energy-cons-diff-1}
    \frac{1}{Re}\int_\Omega \bm{u}\cdot \Delta \bm{v}\,\text{d}\bm{x} = -\frac{1}{Re}\int_\Omega (\nabla \times \bm{u})\cdot (\nabla \times \bm{v})\,\text{d}\bm{x},
\end{equation}
given $\nabla \cdot \bm{u}=0$. Similarly, in \eqref{eq:energy-cons-deriv-pe-1}, 
\begin{equation}\label{eq:energy-cons-diff-2}
    \frac{1}{RePr} \int_\Omega \rho\Delta\rho\,\text{d}\bm{x} = -\frac{1}{RePr}\int_\Omega\nabla \rho \cdot \nabla \rho\,\text{d}\bm{x}.
\end{equation}
Substituting \eqref{eq:energy-cons-time}, \eqref{eq:energy-cons-advec-1}, and \eqref{eq:energy-cons-diff-1} into \eqref{eq:energy-cons-deriv-1}, kinetic energy conservation becomes
\begin{equation}\label{eq:energy-cons-deriv-ke-subs}
    \frac{1}{2}\frac{d}{dt} \int_\Omega \bm{u} \cdot \bm{v} + \frac{1}{Fr}\rho\hat{\bm{z}}\cdot\bm{u}\,\text{d}\bm{x} = -\frac{1}{Re} \int_\Omega (\nabla \times \bm{u})\cdot (\nabla \times \bm{v})\,\text{d}\bm{x}.
\end{equation}
Substituting \eqref{eq:energy-cons-advec-1} and \eqref{eq:energy-cons-diff-2} into \eqref{eq:energy-cons-deriv-pe-1}, the potential energy equation becomes
\begin{equation}\label{eq:energy-cons-deriv-pe-subs}
 \frac{1}{2}\frac{d}{dt}\int_\Omega \rho\cdot \rho - \frac{1}{Fr}\rho\hat{\bm{z}}\cdot\bm{u} \,\text{d}\bm{x}= -\frac{1}{RePr} \int_\Omega (\nabla \rho \cdot \nabla \rho)\,\text{d}\bm{x}.
\end{equation}
Adding together \eqref{eq:energy-cons-deriv-ke-subs} and \eqref{eq:energy-cons-deriv-pe-subs} yields exactly \eqref{eq:total-energy-conservation}. 
\end{appen}

\bibliographystyle{jfm}
\bibliography{jfm}

\begin{thebibliography}{66}
\expandafter\ifx\csname natexlab\endcsname\relax\def\natexlab#1{#1}\fi
\def\au#1{#1} \def\ed#1{#1} \def\yr#1{#1}\def\at#1{#1}\def\jt#1{\textit{#1}} \def\bt#1{#1}\def\bvol#1{\textbf{#1}} \def\vol#1{#1} \def\pg#1{#1} \def\publ#1{#1}\def\arxiv#1{#1}\def\org#1{#1}\def\st#1{\textit{#1}}

\bibitem[Aizinger {\em et~al.\/}(2015)Aizinger, Korn, Giorgetta \& Reich]{aizinger2015}
{\sc \au{Aizinger, Vadym}, \au{Korn, Peter}, \au{Giorgetta, Marco} \& \au{Reich, Sebastian}} \yr{2015}  \at{Large-scale turbulence modelling via $\alpha$-regularisation for atmospheric simulations}.  \jt{Journal of Turbulence}  \bvol{16}~(4),  \pg{367--391}.

\bibitem[Andrews {\em et~al.\/}(2023)Andrews, Shipton \& Wingate]{andrews2023}
{\sc \au{Andrews, Timothy~C}, \au{Shipton, Jemma} \& \au{Wingate, Beth~A}} \yr{2023}  \at{The effect of linear dispersive errors on nonlinear timestepping accuracy in the f-plane rotating shallow water equations}.  \jt{arXiv preprint arXiv:2305.06685} .

\bibitem[Anstey \& Zanna(2017)]{anstey2017deformation}
{\sc \au{Anstey, James~A} \& \au{Zanna, Laure}} \yr{2017}  \at{A deformation-based parametrization of ocean mesoscale eddy {R}eynolds stresses}.  \jt{Ocean Modelling}  \bvol{112},  \pg{99--111}.

\bibitem[Babin {\em et~al.\/}(1995)Babin, Mahalov \& Nicolaenko]{babin1995}
{\sc \au{Babin, A}, \au{Mahalov, A} \& \au{Nicolaenko, B}} \yr{1995} Long-time averaged {E}uler and {N}avier-{S}tokes equations for rotating fluids.  \bt{In {\em Structure and Dynamics of non-linear waves in Fluids, 1994 IUTAM Conference, K. Kirehg{\"a}ssner and A. Mielke (eds), World Scientific\/}},  \pg{pp. 145--157}. World Scientific.

\bibitem[Babin {\em et~al.\/}(1996{\natexlab{{\em a\/}}})Babin, Mahalov \& Nicolaenko]{babin1996}
{\sc \au{Babin, Anatoli}, \au{Mahalov, Alex} \& \au{Nicolaenko, Basil}} \yr{1996{\natexlab{{\em a\/}}}}  \at{Global splitting, integrability and regularity of 3{D} {E}uler and {N}avier-{S}tokes equations for uniformly rotating fluids}.  \jt{European Journal of Mechanics, B/Fluids}  \bvol{15}~(3),  \pg{291--300}.

\bibitem[Babin {\em et~al.\/}(1997{\natexlab{{\em a\/}}})Babin, Mahalov \& Nicolaenko]{babin1997a}
{\sc \au{Babin, Anatoli}, \au{Mahalov, Alex} \& \au{Nicolaenko, Basil}} \yr{1997{\natexlab{{\em a\/}}}}  \at{Regularity and integrability of 3{D} {E}uler and {N}avier--{S}tokes equations for rotating fluids}.  \jt{Asymptotic Analysis}  \bvol{15}~(2),  \pg{103--150}.

\bibitem[Babin {\em et~al.\/}(1999)Babin, Mahalov \& Nicolaenko]{babin1999}
{\sc \au{Babin, Anatoli}, \au{Mahalov, Alex} \& \au{Nicolaenko, Basil}} \yr{1999}  \at{Global regularity of {3D} rotating {Navier-Stokes} equations for resonant domains}.  \jt{Indiana University Mathematics Journal}  \pg{pp. 1133--1176}.

\bibitem[Babin {\em et~al.\/}(2002)Babin, Mahalov \& Nicolaenko]{babin2002}
{\sc \au{Babin, A}, \au{Mahalov, A} \& \au{Nicolaenko, B}} \yr{2002}  \at{Fast singular oscillating limits of stably-stratified 3{D} {E}uler and {N}avier--{S}tokes equations and ageostrophic wave fronts}.  \jt{Large-scale atmosphere-ocean dynamics}  \bvol{1},  \pg{126--201}.

\bibitem[Babin {\em et~al.\/}(1997{\natexlab{{\em b\/}}})Babin, Mahalov, Nicolaenko \& Zhou]{babin1997}
{\sc \au{Babin, Anatoli}, \au{Mahalov, Alex}, \au{Nicolaenko, Basil} \& \au{Zhou, Ye}} \yr{1997{\natexlab{{\em b\/}}}}  \at{On the asymptotic regimes and the strongly stratified limit of rotating {B}oussinesq equations}.  \jt{Theoretical and Computational Fluid Dynamics}  \bvol{9},  \pg{223--251}.

\bibitem[Babin {\em et~al.\/}(1996{\natexlab{{\em b\/}}})Babin, Mahalov \& Nicolaenko]{babin1996b}
{\sc \au{Babin, Anatoli~V}, \au{Mahalov, Alex} \& \au{Nicolaenko, Basil}} \yr{1996{\natexlab{{\em b\/}}}}  \at{Resonances and regularity for boussinesq equations}.  \jt{Russian Journal of Mathematical Physics}  \bvol{4}~(4),  \pg{417--428}.

\bibitem[Bachman {\em et~al.\/}(2018)Bachman, Anstey \& Zanna]{bachman2018}
{\sc \au{Bachman, Scott~D}, \au{Anstey, James~A} \& \au{Zanna, Laure}} \yr{2018}  \at{The relationship between a deformation-based eddy parameterization and the {LANS}-$\alpha$ turbulence model}.  \jt{Ocean Modelling}  \bvol{126},  \pg{56--62}.

\bibitem[Bartello(1995)]{bartello1995}
{\sc \au{Bartello, Peter}} \yr{1995}  \at{Geostrophic adjustment and inverse cascades in rotating stratified turbulence.}  \jt{Journal of the Atmospheric Sciences}  \bvol{52}~(24).

\bibitem[Bennis {\em et~al.\/}(2021)Bennis, Adong, Boutet \& Dumas]{bennis2021lans}
{\sc \au{Bennis, A-C}, \au{Adong, Freddy}, \au{Boutet, Martial} \& \au{Dumas, Franck}} \yr{2021}  \at{{LANS}-$\alpha$ turbulence modeling for coastal sea: {A}n application to {A}lderney {R}ace}.  \jt{Journal of Computational Physics}  \bvol{432},  \pg{110155}.

\bibitem[Burns {\em et~al.\/}(2020)Burns, Vasil, Oishi, Lecoanet \& Brown]{burns2020}
{\sc \au{Burns, Keaton~J}, \au{Vasil, Geoffrey~M}, \au{Oishi, Jeffrey~S}, \au{Lecoanet, Daniel} \& \au{Brown, Benjamin~P}} \yr{2020}  \at{Dedalus: A flexible framework for numerical simulations with spectral methods}.  \jt{Physical Review Research}  \bvol{2}~(2),  \pg{023068}.

\bibitem[Charney(1948)]{charney1948}
{\sc \au{Charney, Jule~G}} \yr{1948}  \at{On the scale of atmospheric motions}.  \bt{In {\em The atmosphere—a challenge: The science of Jule Gregory Charney\/}},  \pg{pp. 251--265}.  \publ{Springer}.

\bibitem[Chen {\em et~al.\/}(1999)Chen, Holm, Margolin \& Zhang]{chen1999direct}
{\sc \au{Chen, Shiyi}, \au{Holm, Darryl~D}, \au{Margolin, Len~G} \& \au{Zhang, Raoyang}} \yr{1999}  \at{Direct numerical simulations of the {N}avier--{S}tokes alpha model}.  \jt{Physica D: Nonlinear Phenomena}  \bvol{133}~(1-4),  \pg{66--83}.

\bibitem[Dukowicz \& Smith(1994)]{dukowicz1994}
{\sc \au{Dukowicz, John~K} \& \au{Smith, Richard~D}} \yr{1994}  \at{Implicit free-surface method for the {B}ryan-{C}ox-{S}emtner ocean model}.  \jt{Journal of Geophysical Research: Oceans}  \bvol{99}~(C4),  \pg{7991--8014}.

\bibitem[Embid \& Majda(1996)]{em1996}
{\sc \au{Embid, Pedro~F.} \& \au{Majda, Andrew~J.}} \yr{1996}  \at{Averaging over fast gravity waves for geophysical flows with arbitary potential vorticity}.  \jt{Communications in Partial Differential Equations}  \bvol{21}~(3-4),  \pg{619--658}.

\bibitem[Embid \& Majda(1998)]{em1998}
{\sc \au{Embid, Pedro~F.} \& \au{Majda, Andrew~J.}} \yr{1998}  \at{Low {Froude} number limiting dynamics for stably stratified flow with small or finite {Rossby} numbers}.  \jt{Geophysical \& Astrophysical Fluid Dynamics}  \bvol{87}~(1-2),  \pg{1--50}.

\bibitem[Farhat {\em et~al.\/}(2014)Farhat, Jolly \& Lunasin]{farhatjollylunasin2014}
{\sc \au{Farhat, Aseel}, \au{Jolly, MS} \& \au{Lunasin, Evelyn}} \yr{2014}  \at{Bounds on energy and enstrophy for the 3{D} {N}avier-{S}tokes-$\alpha$ and {L}eray-$\alpha$ models}.  \jt{Communications on Pure and Applied Analysis}  \bvol{13}~(5),  \pg{2127--2140}.

\bibitem[Foias {\em et~al.\/}(2001)Foias, Holm \& Titi]{foias2001}
{\sc \au{Foias, Ciprian}, \au{Holm, Darryl~D} \& \au{Titi, Edriss~S}} \yr{2001}  \at{The {N}avier--{S}tokes-alpha model of fluid turbulence}.  \jt{Physica D: Nonlinear Phenomena}  \bvol{152},  \pg{505--519}.

\bibitem[Geurts \& Holm(2006)]{geurts-holm}
{\sc \au{Geurts, Bernard~J} \& \au{Holm, Darryl~D}} \yr{2006}  \at{Leray and {LANS}-$\alpha$ modelling of turbulent mixing}.  \jt{Journal of Turbulence} ~(7),  \pg{N10}.

\bibitem[Gjaja \& Holm(1996)]{gjajaholm1996}
{\sc \au{Gjaja, I} \& \au{Holm, DD}} \yr{1996}  \at{Self-consistent wave-mean flow interaction dynamics and its {H}amiltonian formulation for a rotating stratified incompressible fluid}.  \jt{Physica D}  \bvol{98},  \pg{343--378}.

\bibitem[Hecht {\em et~al.\/}(2008{\natexlab{{\em a\/}}})Hecht, Holm, Petersen \& Wingate]{hecht2008lans}
{\sc \au{Hecht, MW}, \au{Holm, DD}, \au{Petersen, MR} \& \au{Wingate, BA}} \yr{2008{\natexlab{{\em a\/}}}}  \at{The {LANS}-$\alpha$ and {L}eray turbulence parameterizations in primitive equation ocean modeling}.  \jt{Journal of Physics A: Mathematical and Theoretical}  \bvol{41}~(34),  \pg{344009}.

\bibitem[Hecht {\em et~al.\/}(2008{\natexlab{{\em b\/}}})Hecht, Holm, Petersen \& Wingate]{hecht2008}
{\sc \au{Hecht, Matthew~W}, \au{Holm, Darryl~D}, \au{Petersen, Mark~R} \& \au{Wingate, Beth~A}} \yr{2008{\natexlab{{\em b\/}}}}  \at{Implementation of the {LANS}-$\alpha$ turbulence model in a primitive equation ocean model}.  \jt{Journal of Computational Physics}  \bvol{227}~(11),  \pg{5691--5716}.

\bibitem[Holm(1999)]{holmfluctuations}
{\sc \au{Holm, Darryl~D.}} \yr{1999}  \at{Fluctuation effects on {3D} {Lagrangian} mean and {Eulerian} mean fluid motion}.  \jt{Physica D: Nonlinear Phenomena}  \bvol{133}~(1--4),  \pg{215--269}.

\bibitem[Holm(2002)]{holm2002karman}
{\sc \au{Holm, Darryl~D}} \yr{2002}  \at{Karman--{H}owarth theorem for the {L}agrangian-averaged {N}avier--{S}tokes--alpha model of turbulence}.  \jt{Journal of Fluid Mechanics}  \bvol{467},  \pg{205--214}.

\bibitem[Holm {\em et~al.\/}(2005)Holm, Jeffery, Kurien, Livescu, Taylor \& Wingate]{holm2005review}
{\sc \au{Holm, Darryl~D}, \au{Jeffery, Chris}, \au{Kurien, Susan}, \au{Livescu, Daniel}, \au{Taylor, Mark~A} \& \au{Wingate, Beth~A}} \yr{2005}  \at{The {LANS}-$\alpha$ model for computing turbulence}.  \jt{Los Alamos Science}  \bvol{29},  \pg{152--171}.

\bibitem[Holm {\em et~al.\/}(2002)Holm, Marsden \& Ratiu]{HolmMarsdenRatiu2002_EPE_GFD}
{\sc \au{Holm, Darryl~D.}, \au{Marsden, Jerrold~E.} \& \au{Ratiu, Tudor~S.}} \yr{2002}  \at{The {E}uler--{P}oincar{\'e} {E}quations in {G}eophysical {F}luid {D}ynamics}.  \bt{In {\em Large-Scale Atmosphere--Ocean Dynamics: Volume II: Geometric Methods and Models\/} (ed. \ed{John Norbury \& Ian Roulstone})},  \pg{pp. 251--300}.  \publ{Cambridge: Cambridge University Press}.

\bibitem[Holm \& Nadiga(2003)]{holmnadiga2003}
{\sc \au{Holm, Darryl~D} \& \au{Nadiga, Balasubramanya~T}} \yr{2003}  \at{Modeling mesoscale turbulence in the barotropic double-gyre circulation}.  \jt{Journal of Physical Oceanography}  \bvol{33}~(11),  \pg{2355--2365}.

\bibitem[Holm \& Titi(2005)]{holmtiti2005}
{\sc \au{Holm, Darryl~D} \& \au{Titi, Ediress~S}} \yr{2005}  \at{Computational models of turbulence: The {LANS}-$\alpha$ model and the role of global analysis}.  \jt{SIAM News}  \bvol{38}~(7),  \pg{1--5}.

\bibitem[Holm \& Wingate(2005)]{holmwingate2005baroclinic}
{\sc \au{Holm, Darryl~D} \& \au{Wingate, Beth~A}} \yr{2005}  \at{Baroclinic instabilities of the two-layer quasigeostrophic alpha model}.  \jt{Journal of Physical Oceanography}  \bvol{35}~(7),  \pg{1287--1296}.

\bibitem[Hoskins {\em et~al.\/}(1985)Hoskins, McIntyre \& Robertson]{hoskins1985}
{\sc \au{Hoskins, Brian~J}, \au{McIntyre, Michael~E} \& \au{Robertson, Andrew~W}} \yr{1985}  \at{On the use and significance of isentropic potential vorticity maps}.  \jt{Quarterly Journal of the Royal Meteorological Society}  \bvol{111}~(470),  \pg{877--946}.

\bibitem[Ilyin {\em et~al.\/}(2006)Ilyin, Lunasin \& Titi]{ilyinlunasintiti2006}
{\sc \au{Ilyin, Alexei~A}, \au{Lunasin, Evelyn~M} \& \au{Titi, Edriss~S}} \yr{2006}  \at{A modified-{L}eray-$\alpha$ subgrid scale model of turbulence}.  \jt{Nonlinearity}  \bvol{19}~(4),  \pg{879}.

\bibitem[Jacobs(1991)]{jacobs1991existence}
{\sc \au{Jacobs, SJ}} \yr{1991}  \at{Existence of a slow manifold in a model system of equations}.  \jt{Journal of Atmospheric Sciences}  \bvol{48}~(7),  \pg{893--902}.

\bibitem[Julien {\em et~al.\/}(2006)Julien, Knobloch, Milliff \& Werne]{julienknobloch2006}
{\sc \au{Julien, Keith}, \au{Knobloch, Edgar}, \au{Milliff, Ralph} \& \au{Werne, Joseph}} \yr{2006}  \at{Generalized quasi-geostrophy for spatially anisotropic rotationally constrained flows}.  \jt{Journal of Fluid Mechanics}  \bvol{555},  \pg{233--274}.

\bibitem[Kim(2018)]{kim2018}
{\sc \au{Kim, Bong-Sik}} \yr{2018}  \at{Attractor dimensions of three-dimensional {N}avier-{S}tokes-$\alpha$ model for fast rotating fluids on generic-period domains: Comparison with {N}avier-{S}tokes equations}.  \jt{arXiv preprint arXiv:1808.09683} .

\bibitem[Kim \& Nicolaenko(2006)]{kim2006}
{\sc \au{Kim, Bong-Sik} \& \au{Nicolaenko, Basil}} \yr{2006}  \at{Existence and continuity of exponential attractors of the three dimensional {N}avier-{S}tokes-$\alpha$ equations for uniformly rotating geophysical fluids}.  \jt{Communications in Mathematical Sciences}  \bvol{4}~(2).

\bibitem[Klainerman \& Majda(1981)]{klainerman1981}
{\sc \au{Klainerman, Sergiu} \& \au{Majda, Andrew}} \yr{1981}  \at{Singular limits of quasilinear hyperbolic systems with large parameters and the incompressible limit of compressible fluids}.  \jt{Communications on Pure and Applied Mathematics}  \bvol{34}~(4),  \pg{481--524}.

\bibitem[Leith(1980)]{leith1980}
{\sc \au{Leith, CE}} \yr{1980}  \at{Nonlinear normal mode initialization and quasi-geostrophic theory}.  \jt{Journal of Atmospheric Sciences}  \bvol{37}~(5),  \pg{958--968}.

\bibitem[Leray(1934)]{leray1934}
{\sc \au{Leray, Jean}} \yr{1934}  \at{Sur le mouvement d'un liquide visqueux emplissant l'espace}.  \jt{Acta mathematica}  \bvol{63},  \pg{193--248}.

\bibitem[Lorenz(1980)]{lorenz1980}
{\sc \au{Lorenz, Edward~N}} \yr{1980}  \at{Attractor sets and quasi-geostrophic equilibrium}.  \jt{Journal of Atmospheric Sciences}  \bvol{37}~(8),  \pg{1685--1699}.

\bibitem[Lorenz(1992)]{lorenz1992slow}
{\sc \au{Lorenz, Edward~N}} \yr{1992}  \at{The slow manifold—what is it?}  \jt{Journal of Atmospheric Sciences}  \bvol{49}~(24),  \pg{2449--2451}.

\bibitem[Lorenz \& Krishnamurthy(1987)]{lorenz1987nonexistence}
{\sc \au{Lorenz, Edward~N} \& \au{Krishnamurthy, V}} \yr{1987}  \at{On the nonexistence of a slow manifold}.  \jt{Journal of Atmospheric Sciences}  \bvol{44}~(20),  \pg{2940--2950}.

\bibitem[Lunasin {\em et~al.\/}(2007)Lunasin, Kurien, Taylor \& Titi]{lunasin2007}
{\sc \au{Lunasin, E}, \au{Kurien, S}, \au{Taylor, MA} \& \au{Titi, ES}} \yr{2007}  \at{A study of the {N}avier--{S}tokes-$\alpha$ model for two-dimensional turbulence}.  \jt{Journal of Turbulence} ~(8),  \pg{N30}.

\bibitem[Lunasin {\em et~al.\/}(2008)Lunasin, Kurien \& Titi]{lunasin2008}
{\sc \au{Lunasin, Evelyn}, \au{Kurien, Susan} \& \au{Titi, Edriss~S}} \yr{2008}  \at{Spectral scaling of the {L}eray-$\alpha$ model for two-dimensional turbulence}.  \jt{Journal of Physics A: Mathematical and Theoretical}  \bvol{41}~(34),  \pg{344014}.

\bibitem[Majda(1984)]{majda1984}
{\sc \au{Majda, Andrew}} \yr{1984} {\em {Compressible Fluid Flow and Systems of Conservation Laws in Several Space Variables}\/},  \st{Applied Mathematical Sciences},  \vol{vol.~53}.  \publ{New York, NY: Springer-Verlag}.

\bibitem[Majda(2003)]{majda-textbook}
{\sc \au{Majda, Andrew}} \yr{2003} {\em Introduction to {PDEs} and Waves for the Atmosphere and Ocean\/},  \st{Courant Lecture Notes in Mathematics},  \vol{vol.~9}.  \publ{American Mathematical Society}.

\bibitem[Majda \& Embid(1998)]{em1998a}
{\sc \au{Majda, Andrew~J.} \& \au{Embid, Pedro}} \yr{1998}  \at{Averaging over fast gravity waves for geophysical flows with unbalanced initial data}.  \jt{Theoretical and Computational Fluid Dynamics}  \bvol{11}~(3),  \pg{155--169}.

\bibitem[Mana \& Zanna(2014)]{portamanazanna2014}
{\sc \au{Mana, PierGianLuca~Porta} \& \au{Zanna, Laure}} \yr{2014}  \at{Toward a stochastic parameterization of ocean mesoscale eddies}.  \jt{Ocean Modelling}  \bvol{79},  \pg{1--20}.

\bibitem[Marsden \& Shkoller(2001)]{marsden2001}
{\sc \au{Marsden, Jerrold~E} \& \au{Shkoller, Steve}} \yr{2001}  \at{Global well--posedness for the {L}agrangian averaged {N}avier--{S}tokes ({LANS}--$\alpha$) equations on bounded domains}.  \jt{Philosophical Transactions of the Royal Society of London. Series A: Mathematical, Physical and Engineering Sciences}  \bvol{359}~(1784),  \pg{1449--1468}.

\bibitem[Petersen {\em et~al.\/}(2008)Petersen, Hecht \& Wingate]{petersen2008}
{\sc \au{Petersen, Mark~R}, \au{Hecht, Matthew~W} \& \au{Wingate, Beth~A}} \yr{2008}  \at{Efficient form of the {LANS}-$\alpha$ turbulence model in a primitive-equation ocean model}.  \jt{Journal of Computational Physics}  \bvol{227}~(11),  \pg{5717--5735}.

\bibitem[Rhines \& Young(1982)]{rhinesyoung1982}
{\sc \au{Rhines, Peter~B} \& \au{Young, William~R}} \yr{1982}  \at{Homogenization of potential vorticity in planetary gyres}.  \jt{Journal of Fluid Mechanics}  \bvol{122},  \pg{347--367}.

\bibitem[Schochet(1994)]{schochet1994}
{\sc \au{Schochet, Steven}} \yr{1994}  \at{Fast singular limits of hyperbolic {PDE}s}.  \jt{Journal of differential equations}  \bvol{114}~(2),  \pg{476--512}.

\bibitem[Shchepetkin \& McWilliams(2005)]{shchepetkin2005}
{\sc \au{Shchepetkin, Alexander~F} \& \au{McWilliams, James~C}} \yr{2005}  \at{The regional oceanic modeling system ({ROMS}): a split-explicit, free-surface, topography-following-coordinate oceanic model}.  \jt{Ocean Modelling}  \bvol{9}~(4),  \pg{347--404}.

\bibitem[Smith \& Lee(2005)]{smithlee2005}
{\sc \au{Smith, Leslie~M} \& \au{Lee, Youngsuk}} \yr{2005}  \at{On near resonances and symmetry breaking in forced rotating flows at moderate {R}ossby number}.  \jt{Journal of Fluid Mechanics}  \bvol{535},  \pg{111--142}.

\bibitem[Smith \& Waleffe(2002)]{smith2002}
{\sc \au{Smith, Leslie~M} \& \au{Waleffe, Fabian}} \yr{2002}  \at{Generation of slow large scales in forced rotating stratified turbulence}.  \jt{Journal of Fluid Mechanics}  \bvol{451},  \pg{145--168}.

\bibitem[Sukhatme \& Smith(2008)]{sukhatmesmith2008}
{\sc \au{Sukhatme, Jai} \& \au{Smith, Leslie~M}} \yr{2008}  \at{Vortical and wave modes in 3{D} rotating stratified flows: random large-scale forcing}.  \jt{Geophysical and Astrophysical Fluid Dynamics}  \bvol{102}~(5),  \pg{437--455}.

\bibitem[Vallis(1996)]{vallis1996}
{\sc \au{Vallis, Geoffrey~K}} \yr{1996}  \at{Potential vorticity inversion and balanced equations of motion for rotating and stratified flows}.  \jt{Quarterly Journal of the Royal Meteorological Society}  \bvol{122}~(529),  \pg{291--322}.

\bibitem[Vanneste \& Yavneh(2004)]{vannesteyavneh2004}
{\sc \au{Vanneste, J} \& \au{Yavneh, I}} \yr{2004}  \at{Exponentially small inertia--gravity waves and the breakdown of quasigeostrophic balance}.  \jt{Journal of the Atmospheric Sciences}  \bvol{61}~(2),  \pg{211--223}.

\bibitem[Ward \& Dewar(2010)]{warddewar2010}
{\sc \au{Ward, Marshall~L} \& \au{Dewar, William~K}} \yr{2010}  \at{Scattering of gravity waves by potential vorticity in a shallow-water fluid}.  \jt{Journal of Fluid Mechanics}  \bvol{663},  \pg{478--506}.

\bibitem[Warn \& Menard(1986)]{warnmenard1986}
{\sc \au{Warn, T} \& \au{Menard, R}} \yr{1986}  \at{Nonlinear balance and gravity-inertial wave saturation in a simple atmospheric model}.  \jt{Tellus A: Dynamic Meteorology and Oceanography}  \bvol{38}~(4),  \pg{285--294}.

\bibitem[Whitehead \& Wingate(2014)]{whiteheadwingate2014}
{\sc \au{Whitehead, Jared~P.} \& \au{Wingate, Beth~A.}} \yr{2014}  \at{The influence of fast waves and fluctuations on the evolution of the dynamics on the slow manifold}.  \jt{Journal of Fluid Mechanics}  \bvol{757},  \pg{155--178}.

\bibitem[Wingate(2004)]{wingate2004}
{\sc \au{Wingate, BA}} \yr{2004}  \at{The maximum allowable time step for the shallow water $\alpha$ model and its relation to time-implicit differencing}.  \jt{Monthly Weather Review}  \bvol{132}~(12),  \pg{2719--2731}.

\bibitem[Wingate {\em et~al.\/}(2011)Wingate, Embid, Holmes-Cerfon \& Taylor]{wingate2011}
{\sc \au{Wingate, Beth~A}, \au{Embid, Pedro}, \au{Holmes-Cerfon, Miranda} \& \au{Taylor, Mark~A}} \yr{2011}  \at{Low {R}ossby limiting dynamics for stably stratified flow with finite {F}roude number}.  \jt{Journal of Fluid Mechanics}  \bvol{676},  \pg{546--571}.

\bibitem[Zhao \& Mohseni(2005)]{zhao2005dynamic}
{\sc \au{Zhao, Hongwu} \& \au{Mohseni, Kamran}} \yr{2005}  \at{A dynamic model for the {L}agrangian-averaged {N}avier-{S}tokes-$\alpha$ equations}.  \jt{Physics of Fluids}  \bvol{17}~(7).

\end{thebibliography}

\end{document}